\documentclass[journal]{IEEEtran}
\usepackage{amssymb}
\usepackage{amsmath}
\usepackage{algorithm}
\usepackage{algorithmic}
\usepackage{multicol}
\usepackage{ragged2e}
\usepackage{subfigure}
\usepackage{lipsum}
\usepackage{mathtools,lipsum}

\usepackage{cuted}
\usepackage{graphicx,mwe}
\usepackage{multirow}
\usepackage{pifont}
\usepackage{rotating}
\usepackage{cite}
\usepackage{xcolor}
\usepackage{authblk}
\begin{document}
	%
	\title{Improved Q-learning based Multi-hop Routing for UAV-Assisted Communication}
	\author{N P~Sharvari,~\IEEEmembership{Student Member,~IEEE,}
		Dibakar~Das,~\IEEEmembership{Senior~Member,~IEEE,}
		Jyotsna~Bapat,~\IEEEmembership{Senior Member,~IEEE,}
		and~Debabrata~Das,~\IEEEmembership{Senior~Member,~IEEE,}}
	\affil{International Institute of Information Technology Bangalore, India.}
	\maketitle
	\begin{abstract} Designing effective Unmanned Aerial Vehicle(UAV)-assisted routing protocols is challenging due to changing topology, limited battery capacity, and the dynamic nature of communication environments. Current protocols prioritize optimizing individual network parameters, overlooking the necessity for a nuanced approach in scenarios with intermittent connectivity, fluctuating signal strength, and varying network densities, ultimately failing to address aerial network requirements comprehensively. This paper proposes a novel, Improved Q-learning-based Multi-hop Routing (IQMR) algorithm for optimal UAV-assisted communication systems. Using \boldmath{$Q(\lambda)$} learning for routing decisions, IQMR substantially enhances energy efficiency and network data throughput. IQMR improves system resilience by prioritizing reliable connectivity and inter-UAV collision avoidance while integrating real-time network status information, all in the absence of predefined UAV path planning, thus ensuring dynamic adaptability to evolving network conditions. The results validate IQMR's adaptability to changing system conditions and superiority over the current techniques. IQMR showcases  36.35\% and 32.05\% improvements in energy efficiency and data throughput over the existing methods.  
		
	\end{abstract}
	
	%
	\IEEEpeerreviewmaketitle
	
	\section{Introduction}
	
	\IEEEPARstart {U}{nmanned} Aerial Vehicles (UAVs) are critical enablers for 5G and beyond networks extending communication capabilities owing to their low cost and ability to provide flexible, adaptable, and high-bandwidth communication coverage in areas where traditional infrastructures are impractical. UAVs, as aerial communication relays, can support a wide range of innovative applications such as environmental sensing, disaster recovery operations, border surveillance, and connectivity for IoT devices \cite{Mozaffari}\cite{surveillance}. Especially in post-disaster scenarios where terrestrial communication systems are destroyed, leaving communities isolated and vulnerable, UAVs provide a rapid and adaptable solution by establishing, maintaining, and restoring communication links, facilitating swift surveillance and transmission of critical information for timely response and support in affected areas \cite{HelpSky}\cite{disaster-area}. 
	
	
	In environments marked by extreme conditions and performance challenges, where continuous end-to-end connectivity cannot be assumed, establishing a reliable data routing mechanism becomes imperative for ensuring successful information dissemination. The possibility of UAVs going out of service due to failures or power constraints and subsequently being replaced by new ones and frequent link disruptions due to the changing positions of UAVs underscores the need for a routing mechanism that accommodates such dynamics \cite{engfly}. In addition to the requirements typical of generic wireless networks, such as finding the most effective route, controlling latency, and ensuring reliability, routing in aerial networks necessitates location awareness, energy considerations, and increased robustness to interconnected links and evolving topologies \cite{paradigm}\cite{mavanet}.
	
	Given the apparent similarity of UAV networks to Mobile Ad Hoc Networks (MANETs) and Vehicular Ad Hoc Networks (VANETs), the MANET and VANET protocols and their variants, are used for potential applications in aerial networks. However, the traditional MANET and VANET protocols may not effectively address the altitude differences, dynamic topologies, energy constraints, line-of-sight challenges, and scalability requirements specific to aerial networks. Thus rendering inadequate for ensuring reliable communication both among UAVs themselves and between UAVs and control center(s) \cite{Gupta}. Q-learning-based UAV-assisted data routing has seen widespread usage in recent times attributed to their intrinsic ability to optimize network path selection by incorporating adaptability and self-optimization, thereby effectively addressing the challenges posed by the dynamic communication environment \cite{RL-based-protocols}\cite{Deep-RL-resource-scheduling}. The existing Q-learning-based routing protocols predominantly prioritize singular parameters such as delay, data transmission velocity, energy constraints, and link quality. However, in scenarios characterized by sporadic end-to-end connectivity and intermittent links, a more nuanced approach considering multiple objectives is imperative. Also, extant research majorly assumes predetermined trajectories for UAVs, neglecting the necessities of unforeseen terrains that may necessitate the real-time adaptive repositioning of UAVs \cite{QL-cognitive-swarm}\cite{multi-agent-routing}. 
	
	In this paper, we explore a scenario involving multiple UAVs tasked with surveying a specific area and relaying the gathered data to the Terrestrial Base Station (TBS). The contributions of the paper include,
	\begin{enumerate}
		\item We propose a novel,  Improved Q-learning-based Multi-hop Routing (IQMR)  algorithm for optimal UAV-assisted communication system by prioritizing energy efficiency, ensuring reliable connectivity, establishing inter-UAV collision avoidance paths, and tracking real-time network status with no predefined UAV path planning. 
		\item We leverage $Q(\lambda)$-learning for routing decisions, enabling adaptation to dynamic environments through continuous learning, thereby enhancing the robustness of the proposed mechanism in unpredictable scenarios. $Q(\lambda)$ improves Q-learning by prioritizing the most relevant experiences for decision-making.  
		\item We introduce a novel objective function for energy-efficient, connectivity and collision-constrained routing to improve network data throughput. Additionally, we devise a novel state and action space framework for individual UAVs, aiming to optimize cumulative reward values across selected source UAV-TBS pairs.
		\item We outline UAV operational modes, including neighbour discovery, transmitting, receiving, and charging, while defining criteria for entering and transitioning between these modes. Incorporating operational modes into the routing mechanism is crucial for coordinating tasks, effectively managing energy consumption, and maintaining network stability.
		\item The results analyze the proposed IQMR's impact on the cumulative system reward for varying parameters, including residual energy, coverage and collision probabilities, packet reception status, and network fragmentation. Comparative evaluations with existing protocols (QMR\cite{QMR} and Q-FANET\cite{QFANET}) showcase IQMR's superior performance in terms of energy consumption efficiency and higher data throughput. 
	\end{enumerate}

	
	
	
	
	
	
	The remainder of the paper is organized as follows. Section II details the related work on protocols designed for UAV data routing, focusing on Q-learning-based protocols. Section III describes the system model of the UAV-enabled cellular communication system for surveillance comprising the descriptions of the multi-UAV network, the Gauss-Markov model for UAV mobility, and the communication channel model. Section IV details the novel Improved Q- Learning based Multi-hop Routing  (IQMR) algorithm and outlines the four modules of IQMR, namely (i) neighbour discovery module, (ii) estimation module for energy-efficient reliable communication, (iii) UAV operational mode configuration module, and (iv) routing decision module. Section V analyses the performance of the proposed IQMR algorithm for changing system parameters and compares IQMR with the existing methods. Finally, the paper is concluded in section VI.
	
	The UAVs are equipped with  Global Positioning System (GPS), cameras, sensors, and wireless communication interfaces to capture and relay information from the surveillance area to the TBS. Employing the $\epsilon$-greedy method, UAVs strike a balance between exploration and exploitation in their decision-making processes. The parameters and symbols used in this paper are listed and explained in Table I.

	\section{Related Work}
	UAV-assisted communication and networking have attracted much interest from both academia and industry. UAVs emerge as prime candidates for surveillance missions due to their capability to effectively monitor remote areas inaccessible to humans. However, to ensure prompt delivery of surveillance data and expedite the information retrieval process, a resilient data forwarding mechanism is essential. This section offers an in-depth exploration of existing UAV routing protocols and their suitability for challenging environments.
	
	In addition to the requirements in conventional wireless networks, such as allowing the network to scale, routing data, meeting latency constraints, and maintaining connectivity and required Quality of Service (QoS), UAV networks should also consider rapidly changing dynamic topology, disappearing nodes, intermittent links, location awareness, and stringent energy constraints. In scenarios where UAV networks operate under adverse conditions, routing protocols must withstand network fluctuations and efficiently deliver information to the ground station with minimal packet loss, message overhead, and delay \cite{DTN-routing} and \cite{LCAD}. Due to the apparent similarity of UAV networks with Mobile Ad Hoc Networks (MANETs) and Vehicular Ad Hoc Networks (VANETs),  the protocols designed for MANETs and VANETs are studied for possible application in aerial networks. However, the existing routing methods lack flexibility and are limited by space and time. Therefore, there is a need for a tailor-made routing approach that suits the application-specific requirements of aerial networks to have reliable communication among the UAVs and from UAVs to the control center(s) \cite{Gupta}. 
	
	The complex and diverse flight environments have caused airborne networks to be in an unpredictable, random state. Therefore, the conventional protocols will not be able to adapt to changing networks in real-time and cater to the needs of the UAV networks. Hence, flexible and autonomous learning-based protocols are envisioned as promising solutions that comply with the unique requirements of UAV networks. Q-learning algorithms have gained significant importance in optimizing data routing for UAVs, thus improving network performance and reliability. These mechanisms utilize reinforcement learning techniques to enable UAVs to make intelligent routing decisions based on learned experiences.
	
	Q-Learning-Based Geographic (QGeo)\cite{QGeo} protocol incorporates packet travel speed to determine the next forwarder based on packet travel time, MAC delay, and transmission delay. Reward Function Learning for QL-Based Geographic Routing Protocol (RFLQGEO)\cite{RFLQGEO} extends QGeo by considering UAV position information obtained through GPS to compute the distance progress of data towards the sink while considering MAC and transmission delays. Q-Learning-Based Cross-Layer Routing Protocol (QLCLRP)\cite{QLCLRP} integrates carrier sense multiple access with multi-channel synchronization to ensure reliable data transmission in UAV networks. Q-Learning-Based Balanced Path Routing (QLBR) \cite{QLBR} is designed to achieve load balancing and efficient resource utilization by dynamically selecting balanced paths based on network congestion, link quality, and available resources. Q-Learning-based Multi-Objective Routing (QMR)\cite{QMR} focuses on optimizing delay and energy consumption in data routing. Q-Learning-based Routing Protocol for FANET  (Q-FANET) \cite{QFANET} aims to enhance the Quality of Service in dynamic UAV networks by minimizing delay and considering channel conditions and residual energy levels. QL-Based Topology-Aware Routing (QTAR) \cite{QTAR} leverages two-hop neighbour information to facilitate efficient data forwarding, considering factors like transmission delay and changing network conditions. However, QGeo, RFLQGEO, QLCLRP, QLBR, and Q-FANET overlook energy efficiency as a routing parameter and lack adaptability to changing network characteristics such as topology, link quality, mobility, and obstacles. Although QMR estimates the link quality using the expected transmission count to update the Q-value, considering the SIR  as a metric would provide a more accurate estimation of the link status. In QTAR, the maintenance of two-hop neighbour information leads to increased network overhead. 
	
	Present Q-learning-based routing protocols often focus on individual factors like delay, data transmission speed, energy limitations, and link quality. However, adopting a more intricate approach that considers multiple network parameters becomes essential in scenarios characterized by irregular end-to-end connectivity, vanishing nodes and finicky links. Additionally, existing research operates under the assumption of predetermined trajectories for UAVs, overlooking the challenges posed by unexpected terrains that may require UAVs' dynamic and real-time repositioning. Hence, there is a requirement for specialized protocols that address the unique needs of airborne networks, offering adaptability to changing topology and demand-specific routing, particularly in scenarios without a predetermined UAV path. Q-learning-based adaptive routing protocols explored lack simultaneous multi-objective optimization of parameters crucial for data routing in dynamic UAV networks. Our research proposes an integrated approach to develop a novel packet forwarding method that incorporates factors such as UAV energy consumption, coverage and collision constraints, and packet reception status without prior trajectory planning. Additionally, our approach dynamically adjusts Q-learning parameters (learning rate and discount factor) and neighbour discovery (incidence of \textit{HelloInterval}) to respond effectively to changing network characteristics. To the best of our knowledge, this proposed novel approach has not been addressed in the existing literature.

\section{System Description}
\subsection{Network Model}
The paper considers a network of low-altitude UAVs that communicate with each other and a Terrestrial Base Station (TBS). The network consists of $M$ UAV nodes denoted as $\mathbb{U} = [U_1,\dots, U_M]$. Each UAV, represented by $U_i$, is characterized by parameters such as residual energy ($E^{res}$), packet reception status ($P^{rs}$), coverage ($P^{cov}$), and collision ($P^{coll}$) probabilities.

The UAVs are deployed within a cylindrical 3D space with a $P$ radius and a maximum $H$ height above the ground. The TBS is at the center of the base of the cylindrical region. UAVs move in discrete time slots using the Gauss Markov Mobility Model (GMMM) while having no predetermined trajectories, all within the confines of a cylindrical 3D space. The location of the UAVS and the TBS are represented as $(x_i,y_i,h_i)  \forall i \in M $ and $(x_T,y_T,h_T)$ respectively, as in Fig.2.

\begin{figure}
	\centering
	\includegraphics[width=6cm, height=6cm]{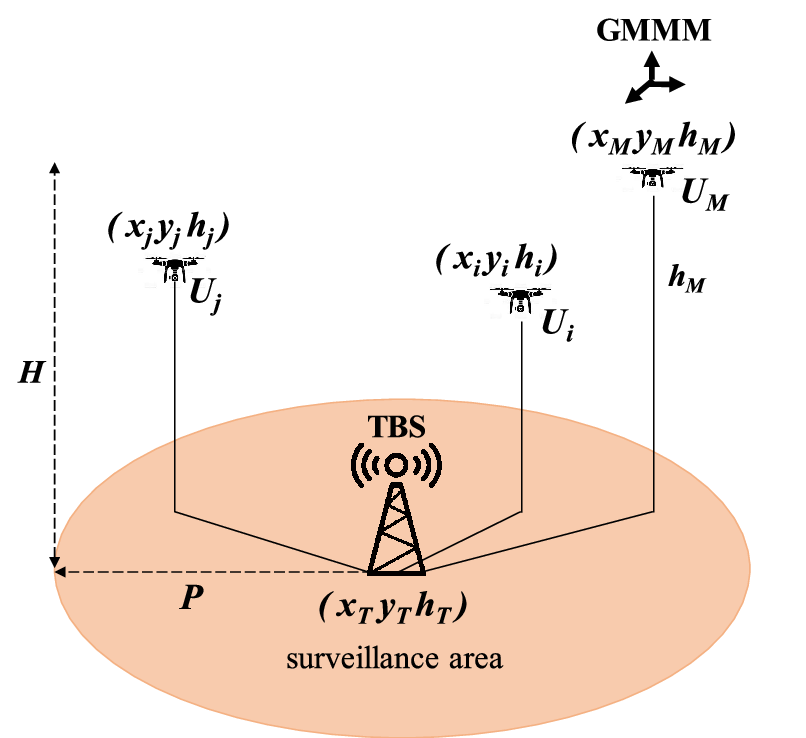}
	\caption{Multi-UAV network with 3D  Gauss Markov Mobility Model.
	}
\end{figure}

\subsection{Gauss Markov Mobility Model}
The UAVs follow the 3D Gauss-Markov Mobility Model to guide their movement process in the air \cite{LADTR}. 
Each UAV is assigned a speed ($s$), direction ($d$), and vertical pitch ($p$) (angle of ascent or descent affecting the UAV's altitude change computed w.r.t the ground plane). The UAV calculates $s$, $d$ and $p$ at every set time $(n)$ using the Gauss-Markov equations as,
\begin{equation}
	s_n = \alpha s_{n-1}+(1-\alpha) \bar{s} + \sqrt{(1-\alpha^2)}s_{x_{n-1}}
\end{equation}
\begin{equation}
	d_n = \alpha d_{n-1}+(1-\alpha) \bar{d} + \sqrt{(1-\alpha^2)}d_{x_{n-1}}
\end{equation}
\begin{equation}
	p_n = \alpha p_{n-1}+(1-\alpha) \bar{p} + \sqrt{(1-\alpha^2)}p_{x_{n-1}}
\end{equation}

The tuning parameter $\alpha$ takes a value within the range of (0,1). The mean speed, direction, and pitch are denoted as $\bar{s}$, $\bar{d}$, and $\bar{p}$, respectively. To introduce randomness to the new speed, direction, and pitch, random variables $s_{x_{n-1}}$, $d_{x_{n-1}}$, and $p_{x_{n-1}}$ are chosen from uniform distributions $[s_{min}, s_{max}]$, $[d_{min}, d_{max}]$, and $[p_{min}, p_{max}]$, respectively. The next position of the UAV ($x_{n+1}, y_{n+1}, h_{n+1}$) is determined from the current location ($x_n, y_n h_n$), the speed ($s_n$), direction ($d_n$), and pitch ($p_n$) as,
\begin{equation}
	x_{n+1} = x_{n}+s_{n}cos(d_{n})cos(p_{n})
\end{equation}
\begin{equation}
	y_{n+1} = y_{n}+s_{n}sin(d_{n})cos(p_{n})
\end{equation}
\begin{equation}
	h_{n+1} = h_{n}+s_{n}sin(p_{n})
\end{equation}
UAVs pause for a time duration $t_p$ (pause time) at each location to relay the collected information towards TBS via multi-hop UAV path.
\begin{table}[h]
	\begin{center}
		\caption{List of  parameters/ symbols}
		\begin{tabular}{p{0.25\columnwidth}|p{0.65\columnwidth}}
			\hline
			\vspace{1pt}\textbf {Parameter/ Symbol} & \vspace{1pt}\textbf{Significance}\\ 
			\hline
			\vspace{1pt} $U_i$ & \vspace{1pt} UAV denoted as $i$, $\forall \hspace{2pt} i \hspace{2pt}\in$ (1 to $M$) \\
			$M$, $N_T$ and $N_C$ & total UAVs in the network, neighbour table set and candidate neighbour set\\
			$r$ &  inter UAV distance\\
			$r_{min}$ and  $D$  & minimum and relative inter UAV distance\\
			$h$, $s$, $d$, and $p$ & height, speed, direction, and pitch of UAV operation\\ 
			$t^{LST}$ & link sustenance time\\
			$E^{res}, P^{rs}, P^{cov}$ and $P^{coll}$ & normalized residual energy, packet reception status, coverage and collision probability\\
			$E_{th}, SIR_{th}, P^{coll}_{th} $ & threshold energy, signal to interference ratio and collision probability\\
			$\mathcal{E}(k,r)$ & transmission energy required to transmit $k$ bits of data over a distance $r$\\
			$PAC^{tx}(L2)$ and $PAC^{tx}(L3)$ & packets transmitted at layer 2 and layer 3\\
			$ACK^{tx}(L2)$ and  $ACK^{tx}(L3)$& acknowledgements received at layer 2 and layer3\\
			$\zeta$ and $g$ & path loss component and channel power fading gain\\
			$R_t$ & uniform transmission range of UAV\\
			$q_r$ and $q_t$ & reception and transmission queue\\
			$S$, $A$ and $R$ & state space, action space and reward\\
			$Q^{old}(.)$ and $Q^{new}(.)$ & old and new network Q-matrix corresponding to the condition $(.)$\\
			$\beta$ and $\gamma$ & dynamic learning rate and discount factor\\
			$\lambda$ & learning rate controlling the eligibility trace w.r.t. $Q(\lambda)$ learning\\
			\hline
		\end{tabular}
	\end{center}
\end{table}

\subsection{Channel Model}
The propagation through the wireless channel is a combination of distance-dependent path loss attenuation and small-scale fading. We adopt the standard power-law path loss model and define the path loss function as $l(r_{ij}, h_i,\zeta )= (r_{ij}^2 + h_i^2)^{-\zeta/2}$ where $r_{ij}$ denotes the horizontal distance between the  transmitting UAV $i$ and  receiving UAV $j$, $h_i$ is the operating height of the  UAV $i$ and $\zeta$ is the path loss exponent \cite{pathloss}. We assume Nakagami-$m$ small-scale fading channel with fading parameter $m$ computed as $m \triangleq 2(K+1)/(2K+1)$ where $K$ is the Rician $K$-factor characterizing the ratio between the powers of direct and scattered paths \cite{nakagami}. Therefore, the channel power fading gain $g_{ij}$ follows the Gamma distribution $\Gamma(m,\frac{1}{m})$.

Assuming interference-limited scenario \cite{nakagami}, we define the Signal to Interference Ratio (SIR) as,
\begin{equation}
	SIR_{ij} \triangleq \frac{g_{ij}l(r_{ij},h_i)}{I}
\end{equation}
$I$ is the aggregate interference power defined as $ I \triangleq \sum_{u\in\mathbb{U}\setminus\{i\}}g_{uj}l(r_{uj},h_u)$.

\subsection{Problem Formulation}
The objective involves employing UAVs to survey a specific area, capture  information and transmit the same to a TBS through a multi-UAV pathway. The overarching goal is to optimize the transmission process by maximizing residual energy, enhancing successful data transmission, improving coverage probability, and minimizing potential collisions. The problem formulation of multi-hop data routing for energy-efficient and reliable communication over the distributed UAV network is as follows, 
\begin{equation}
	max({E}_n^{res}(i)), max({P}_n^{rs}(i)), max({P}_n^{cov}(i)), min({P}_n^{coll}(i))
\end{equation}
subject to:
\begin{equation}
	{E}_n^{res}(i) \geq E_{th} 
\end{equation}

\begin{equation}
	{P}_n^{cov}(i)=\mathbb{E}[SIR_n(i)\geq SIR_{th}] 
\end{equation}
\begin{equation}
	{P}_n^{coll}(i) \leq P^{coll}_{th} \hspace{2pt} \& \hspace{2pt} r \geq r_{min}\hspace{0.2cm}
\end{equation}
\begin{equation*}
	\forall \hspace{2pt}n \in N, \forall \hspace{2pt} i \in M
\end{equation*}
The objective function in (8)  includes residual energy ($E^{res}$), packet reception status ($P^{rs}$), coverage probability($P^{cov}$), and collision probability ($P^{coll}$). Residual energy signifies the remaining UAV power. 	Assuming a three-layer network architecture  (physical layer, MAC layer and network layer), we account for the successful transmission of packets at both layers 2 and 3 by considering the status of packet reception. Packet reception status includes transmitted packets and acknowledgements at Layers 2 and 3. Layer 2 guarantees local communication with assured inter-node packet delivery. Layer 3 oversees end-to-end routing, ensuring packet transfer from the source UAV to the TBS through a multi-node path. Coverage probability sets the minimum SIR for reliable UAV communication. Collision probability indicates the likelihood of two dynamically maneuvering  UAVs colliding in a given airspace. Equation (8) considers the normalized parameters. The normalization of the parameters follows the below relation, 
\begin{equation}
	\hat{x}=\frac{x_{actual}-x_{minimum}}{x_{maximum}-x_{minimum}}
\end{equation}
In (12), $x$ indicates the parameter to be normalized. The network runs for $N$ number of episodes. The equations from (9) to (11) state the constraints for the objective (8). Condition (9) specifies the energy constraint, requiring the residual energy to be greater than the defined threshold ($E_{th}$). Constraint (10) establishes the reliability criteria, necessitating a signal-to-interference ratio equal to or greater than the predefined threshold ($SIR_{th}$). Constraint (11) ensures both collision probability and minimum inter-UAV distance requirements are met; otherwise, requiring the UAVs to relocate randomly. The collision probability must be less than or equal to the threshold ($P^{coll}_{th}$), and the inter-UAV distance ($r$) must be greater than or equal to a minimum value ($r_{min}$) to avert physical contact. The paper's next section enumerates the solution for the optimization problem in (8).

\section{Improved Q-learning-based Multi-hop Routing Algorithm for Energy Efficient Reliable Communication}

This section presents a novel, Improved Q-learning-based Multi-hop Routing (IQMR) algorithm, addressing the recognized demand for energy-efficient, reliable communication with no inter-UAV collisions in multi-hop routing scenarios. Illustrated in Fig. 3, the block diagram outlines the components of the proposed IQMR algorithm, which includes four modules: (i) neighbour discovery module, (ii) estimation module for energy-efficient reliable communication, (iii) UAV operational mode configuration module, and (iv) routing decision module. \textit{Algorithm 2} elucidates the procedural steps of the proposed IQMR mechanism. 
\begin{figure}
	\centering
	\includegraphics[width=9cm, height=10cm]{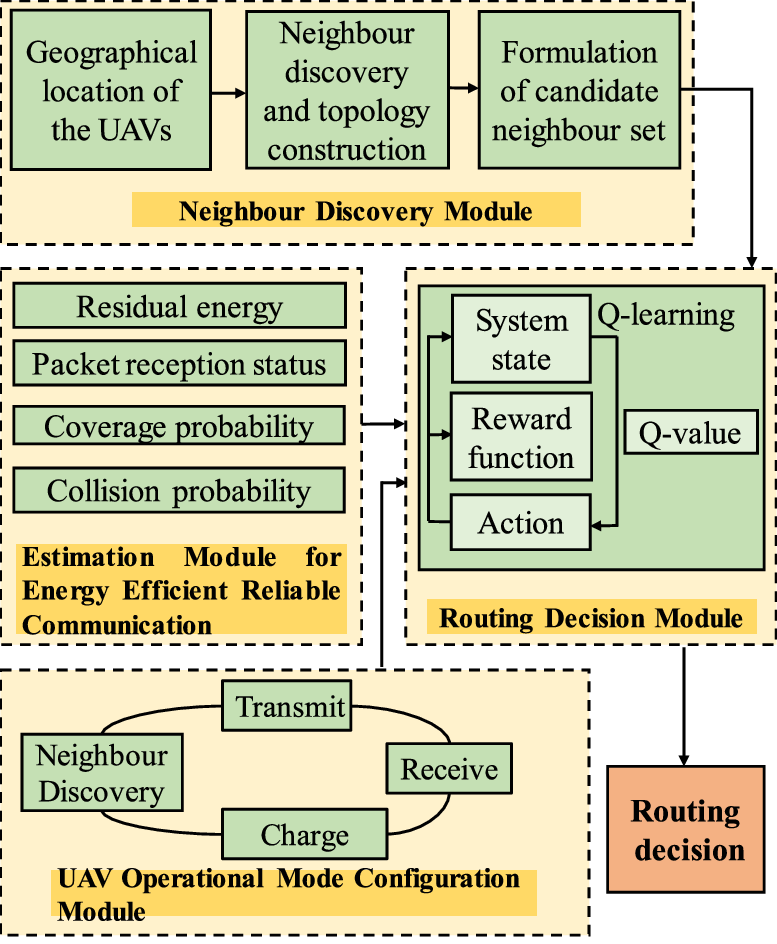}
	\caption{Framework of Improved Q-learning-based Multi-hop Routing (IQMR) Algorithm for Energy Efficient Reliable Communication.}
\end{figure}

\subsection{Neighbour Discovery Module}
The neighbour discovery module, described in \textit{Algorithm 1},  constructs and updates the network topology by maintaining a  \textit{NeighbourTable} ($N_T$). Neighbour discovery involves exchanging \textit{HelloMessage} over every \textit{HelloInterval}. Each \textit{HelloMessage} includes UAV location, residual energy, packet reception status, learning rate, discount factor, and Q-value. The receipt of the \textit{HelloMessage} establishes the UAV's neighbour set, consisting of the nodes from which the UAV receives the message. A subset of neighbour UAVs, confined to a sector $\mathbb{S}$, are considered potential candidate neighbours $(N_C)$. The sector $\mathbb{S}$ is chosen, with a radius of $R_t$ and an angle of $\pi/2$ around the axis connecting the source node (transmitting UAV) and the destination node (TBS) for the targeted progression of data towards the intended destination (TBS), as depicted in Fig. 4 \cite{my}. The \textit{NeighbourTable} stores information retrieved from the \textit{HelloMessage} sent by candidate neighbours.

Determining the appropriate timing for the \textit{HelloInterval} is crucial to adapt to the dynamically changing network. The \textit{HelloInterval} should be set to a frequency allowing timely updates of the \textit{NeighbourTable} while minimizing unnecessary overhead. By evaluating the Link Sustenance Time ($t^{LST}$), we can estimate the incidence of \textit{HelloInterval} based on the acquired topology information. The $t^{LST}$  represents the duration the link between communicating UAVs remains stable and reliable, ensuring collision-free uninterrupted connectivity.

The estimation of $t^{LST}$ is influenced by the relative motion between the two communicating UAVs. The three scenarios for determining $t^{LST}$ include UAVs (i) receding, (ii) approaching, and  (iii)  moving equidistant apart. Let $U_i$ and $U_j$ be the two communicating UAVs, and $R_t$ be the uniform transmission range of each UAV. The relative distance ($D_{ij}(n)$) between the two UAVs $U_i$ and $U_j$ at time instant $n$ is determined using the Euclidean distance formula as
\begin{multline}
	D_{ij}(n) =\\ \sqrt{[x_i(n)-x_j(n)]^2 + [y_i(n)-y_j(n)]^2 + [h_i(n)-h_j(n)]^2}	
\end{multline}
Assume UAVs $U_i$ and $U_j$ are moving at speeds $s_i$ and $s_j$, respectively.  When the UAVs are receding, the expected time for their separation to exceed their transmission range ($R_t$) is  $t^{LST}_{ij}(n) = [2R_t- D_{ij}(n)]/[s_i+s_j]$. Conversely, when the UAVs are approaching,  the expected time for their separation to equal the minimum permissible inter-UAV distance ($r_{min}$) is $t^{LST}_{ij}(n)= [D_{ij}(n)-r_{min}/|s_i-s_j|]$. Adjusting the next occurrence of the \textit{HelloInterval} as $H_I(next) = H_I(n) + t^{LST}_{ij}(n)$ enables UAVs to sustain communication in a dynamic network environment while mitigating the risk of collision, by accounting for the estimated time needed for separation to either exceed the $R_t$ in receding scenarios or reach $r_{min}$ when UAVs are approaching. In the case where UAVs are moving equidistantly apart,   their relative positions remain unchanged over time, causing the \textit{HelloInterval} to repeat at each time instant $n$ consistently. Suppose the UAVs fail to receive the \textit{HelloMessage}  associated with an existing record within a specific expiration time of \textit{HelloInterval} (set to $300$ ms); the corresponding entry is removed from the \textit{NeighbourTable}.
\begin{figure}
	\centering
	\includegraphics [width=0.75\linewidth] {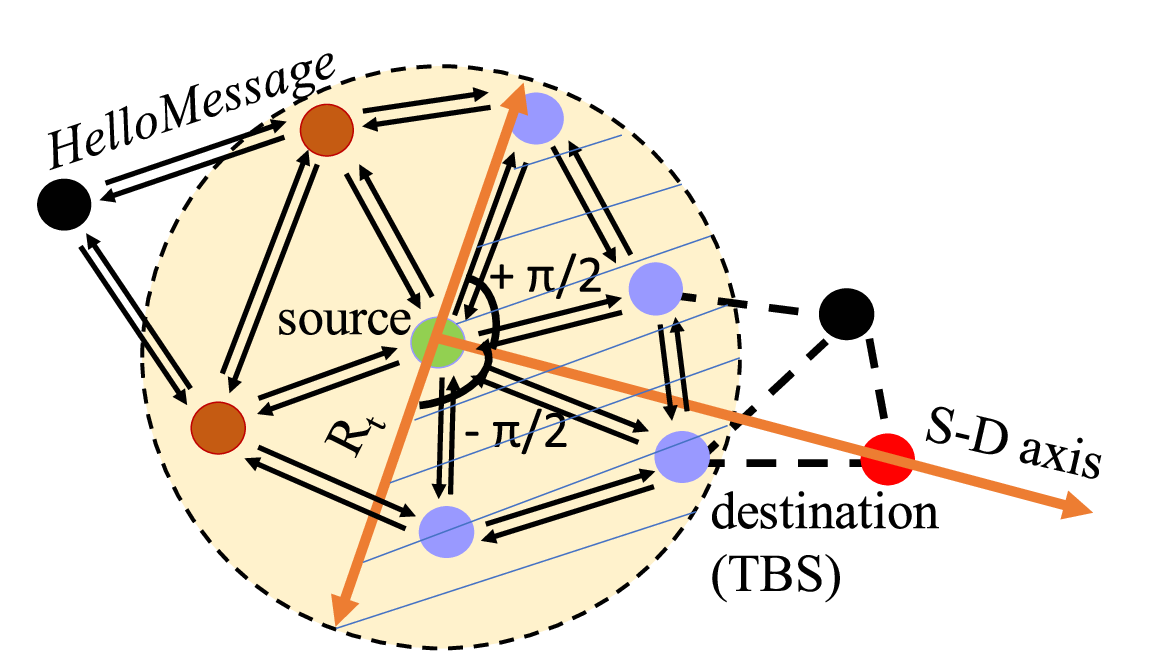}
	\caption{Representation of sector $\mathbb{S}$  (shaded region)  and formulation of candidate neighbour set (nodes within $\mathbb{S}$). Representation of nodes are as black-non neighbours, brown-neighbours, blue-candidate neighbours, green-source, red-destination.}
\end{figure}

\begin{algorithm}
	\caption{: Hello Message Exchange for Topology Construction}\label{algo1}
	\textbf{Input:} \\
	Location of ${M}$ UAVs and TBS, \textit{HelloMessage} and sector $\mathbb{S}(R_t,\pi)$ . \\
	\textbf{Output:} Updated \textit{NeighbourTable} ($N_T$).\\
	\textbf{Initialization:}\\
	Set UAV identification number as $U_i = [U_1,\ldots, U_M] $\\
		\raggedright{\textbf{Phase 1: Broadcast \textit{HelloMessage}}}\\
		\begin{algorithmic}[1]
			\FOR{each UAV $U$ in $U_1$ to $U_M$}
			\STATE broadcast \textit{HelloMessage}
			\ENDFOR
		\end{algorithmic}
		\raggedright{\textbf{Phase 2: \textit{NeighbourTable} construction}}\\
		\begin{algorithmic}[1]
			\FOR{each received \textit{HelloMessage}}
			\STATE get originator UAV id $U_i$  \COMMENT{id of UAV broadcasting \textit{HelloMessage}}
			\STATE get originator location $Location_{U_i}$
			\IF{$Location_{U_i} \in \mathbb{S}$} 
			\STATE Add originator to Candidate Neighbour Set {$N_C \leftarrow U_i$} 
			\IF{$U_i\in N_T$} 
			\STATE update existing record of $U_i$
			\STATE remove existing record of $U_j$\hspace{2pt}$\in N_T$,  $U_j$\hspace{2pt}$\neq U_i$ 
			\ELSE
			\STATE add a new record for $U_i$
			\STATE remove existing record of $U_j$\hspace{2pt}$\in N_T$,  $U_j$\hspace{2pt}$\neq U_i$ 
			\ENDIF
			\ENDIF
			\ENDFOR
			\STATE update \it {HelloInterval}
		\end{algorithmic}
	\end{algorithm}
	\subsection{Estimation Model for  Energy Efficient Reliable Communication}
	This section presents estimation models for residual energy, packet reception status, coverage and collision probabilities. 
	\subsubsection{Residual Energy Estimation Model}
	The proposed work employs a distance-dependent energy consumption model  for estimating the energy expended for data transmission \cite{energy}. Suppose the distance ($r$) between the transmitting UAV and the receiving UAV is less than a certain threshold ($r_0$); we use a free space channel model ($d^2$ power loss); otherwise,  multi-path fading model ($d^4$ power loss). Thus the expression for energy expended to transmit $k$ bits of data over a distance $r$ is \cite{eng-eqs},
	\begin{equation}
		{E}^{tx}(k,r)=
		\begin{cases}
			\epsilon_{elec}k + \epsilon_{amp-fs}kr^2 \hspace{1.5cm} r \leq r_0\\
			\epsilon_{elec}k + \epsilon_{amp-mp}kr^4 \hspace{1.5cm} r>r_0
		\end{cases}
	\end{equation}
	where $\epsilon_{elec}$ and $\epsilon_{amp}$ are system constants. $\epsilon_{elec}$ is the energy dissipated by the radio to run the transmitter-to-receiver circuitry (J/bit), $\epsilon_{amp-fs}$ is the distance-dependent amplifier energy with free space channel model (J/bit/m$^2$) and $\epsilon_{amp-mp}$ is the distance-dependent amplifier energy with multi-path channel model (J/bit/m$^4$). 
	
	In addition to the energy expended during data transmission, the UAV also consumes energy for its flight. The UAV logs the depleted battery capacity as it is equipped with a battery management system. Then, the model verifies if the remaining energy ($E^{res}n(i)$) of each UAV $i$ at time instant $n$ is greater than or equal to the specified threshold ($E_{th}$).
	\subsubsection{Packet Reception Status}
	Packet reception status provides real-time feedback on the successful receipt of packets. It includes transmitted data packets at layer 2 ($PAC^{tx}(L2)$) and layer 3 ($PAC^{tx}(L3)$) and acknowledgements at layer 2 ($ACK(L2)$) and layer 3 ($ACK(L3)$). The packet reception status during each episode of data transmission is determined by the reception status at Layer 2 and Layer 3, respectively, assuming that certain packets are always transmitted ($PAC^{tx}(L2) \neq 0, PAC^{tx}(L3)\neq 0$).
	\begin{equation}
		{P}_n^{rs}(i)=  \sum_{n=0}^{N} {P}_n^{rs}(L2)(i) + {P}_n^{rs}(L3)(i) \hspace{0.1cm}\forall\hspace{0.1cm} i \in M
	\end{equation}
	\begin{multline}
		{P}_n^{rs}(i)= \sum_{n=0}^{N}  \frac{ACK_n(L2)(i)}{PAC^{tx}_n(L2)(i)} + \frac{ACK_n(L3)(i)}{PAC^{tx}_n(L3)(i)} \hspace{0.1cm} \forall\hspace{0.1cm} i \in M
	\end{multline}
	In (16), the expressions $PAC^{tx}_n(L2)(i)$ and $PAC^{tx}_n(L3)(i)$ denote the cumulative number of Layer 2 and Layer 3 packets transmitted, respectively, by the  UAV $i$ up to the $n^{th}$ episode of data transfer. Similarly, $ACK_n(L2)(i)$ and $ACK_n(L3)(i)$ represent the cumulative count of Layer 2 and Layer 3 acknowledgements received by the  UAV $i$ corresponding to the packets it has transmitted. These parameters track the ongoing data transfer process.
	\subsubsection{Coverage Probability}
	The coverage probability represents the minimum SIR requirement between communicating UAVs for reliable data transfer. Following the analytical model in \cite{my} ${P}^{cov}_n(i)$ for  UAV $i$ at  time instant $n$ is  as follows.
	\begin{equation}
		{P}^{cov}_n(i) = \mathbb{E}[SIR_n(i)\geq SIR_{th}]
	\end{equation}
	where $SIR_{th}$ is the minimum SIR requirement.
	\subsubsection{Collision Probability}
	The collision probability indicates the likelihood of two UAVs colliding when their separation is below the minimum required threshold ($r_{min}$) to avoid a collision. Based on the mathematical model in \cite{my}, the expression for the collision probability ${P}^{coll}_n(i)$ of the  UAV $i$ at  time instant $n$ is as follows. 
	\begin{equation}
		{P}^{coll}_n(i) = 1- exp\bigg(\frac{-r^2_n(i)}{2 \xi^x_n(i) \xi^y_n(i)}\bigg)  
	\end{equation}
	$r$ is the radius of the area swept by the two approaching UAVs; the "area swept by the two approaching UAVs", also called collision cross-sectional area, refers to the combined space covered by the paths of two UAVs as they move toward each other. The positions of the approaching UAVs are assumed such that one lies at the center and the other on the circumference of the collision cross-sectional area; thus, the radius ($r$) is considered as the inter-UAV distance. $\xi^x$ and $\xi^y$ are divergence in the UAV trajectory in $x$ and $y$ directions respectively. 
	\subsection{UAV Operational Mode Configuration Module}
	The operational modes of UAVs represent the distinct phases in their functioning, including \textit{NeighbourDiscovery}, \textit{Receive}, \textit{Transmit}, and \textit{Charge}. Fig. 5 illustrates the sequential transition between the modes. UAVs with energy levels greater than the specific threshold (${E}^{res} > E_{th}$) operate in \textit{NeighbourDiscovery}, \textit{Receive} and \textit{Transmit} modes. Otherwise, they transition to the \textit{Charge} mode, relocating to designated ground charging points to replenish energy before rejoining the network in \textit{NeighbourDiscovery} mode. The UAVs start with \textit{NeighbourDiscovery} until the \textit{HelloInterval} duration expires. Subsequently, they transition to the \textit{Receive} mode, where the data is received and moved to the reception queue ($q_r$). Depending on the nature of the received data, whether it is surveillance data, \textit{ACKs}, or \textit{HelloMessage}, specific actions are performed, such as moving data to the transmission queue ($q_t$), adjusting the reward function in equation (19), or updating the \textit{NeighbourTable} (\textit{Algorithm 1}). UAVs enter the \textit{Transmit} mode only when there is data to relay ($q_t \neq 0$). In \textit{Transmit} mode, UAVs evaluate ${E}^{res}$ in (14), ${P}^{rs}$ in (16), ${P}^{cov}$ in (17) and ${P}^{coll}$ in (18) and update the reward function to facilitate optimal next-hop node selection. \textit{NeighbourDiscovery} mode recommences either at the beginning of the \textit{HelloInterval} or when the UAV is fragmented ($N_C = 0$) from the network (when a UAV has no candidate neighbours, it indicates that it is fragmented from the network, and to restore connectivity, the UAV continuously broadcasts \textit{HelloMessages} to notify other UAVs of its presence).
	
	\begin{figure}
		\centering
		\includegraphics[width=0.49\textwidth]{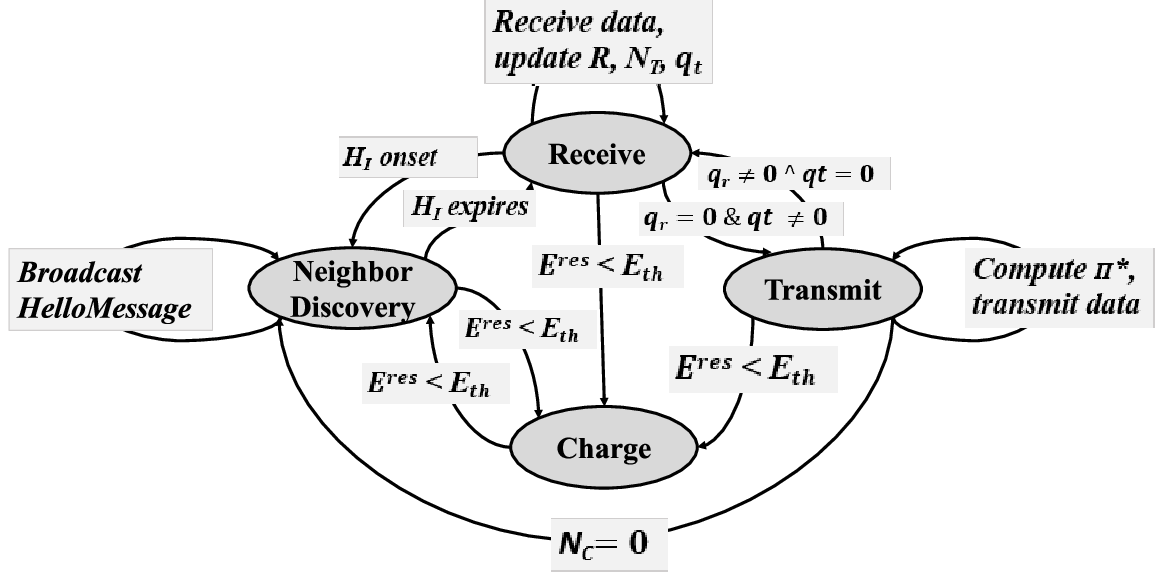}
		\caption{Operating modes of UAV.}
	\end{figure}
	
	\subsection{Routing Decision Module}
	This section proposes a Routing Decision Module that employs an $Q(\lambda)$-learning framework to determine the appropriate next-hop node for transmitting surveillance data to the TBS through a multi-hop path. The module leverages information from the Neighbour Discovery Module, the Estimation Module, and the UAV Operational Mode Configuration Module. 
	
	\subsubsection{Q-Learning based Multi-hop Routing}
	The proposed $Q(\lambda)$-learning-based multi-hop routing integrates operational mode and state information to enhance the routing algorithm. The UAV operating modes are neighbour discovery, receive, transmit, and charge. The UAV states include residual energy, packet reception status, coverage and collision probabilities. Based on the operational mode and state parameters, an approximate action towards data forwarding, which leads to maximized immediate reward value, is chosen. Over time, the UAVs learn the best next-hop node to forward the data that maximizes the cumulative reward values between the source UAV and the TBS. For $Q(\lambda)$-learning, we define a tuple $(S, A, R)$ where $S$ represents the state space, $A$ indicates the action space and $R$ specifies the reward space. Below is a detailed description of the state, action, and reward space.
	\paragraph{States}
	At instant $n$, the network state is given as $	\boldsymbol{S}_n = [\boldsymbol{E}_n^{res}(i), \boldsymbol{P}_n^{rs}(i), \boldsymbol{P}_n^{cov}(i), \boldsymbol{P}_n^{coll}(i) ]$ The vectors $\boldsymbol{E}_n^{res}(i)$, $\boldsymbol{P}_n^{rs}(i)$, $\boldsymbol{P}_n^{cov}(i)$ and $\boldsymbol{P}_n^{coll}(i)$ represent the residual energy, packet reception status, coverage, and collision probabilities in their normalized form for the set of $M$ UAVs ($\forall \hspace{2 pt} i \in M$). 
	
	\paragraph{Actions}
	The action space, denoted as $\boldsymbol{A}_n$, is defined as $\boldsymbol{A}_n = \boldsymbol{a}_n(i)$, where $\boldsymbol{a}_n(i)$ represents the routing decision of the UAV $i$ at  discrete time $n$. This decision encompasses  the choice of the next-hop node. 
	\paragraph{Rewards}
	The reward function maps states to actions by assigning positive reward values corresponding to the action taken. These rewards define the effectiveness of successful data transmission in a dynamic network. Thus, the reward function corresponding  to the action is defined as a joint metric of normalized  parameters in the objective function (8) as,
	
	\begin{equation}
		R(s_n,a_n)=
		\begin{cases}
			w_1(1-{P}_n^{coll})+ w_2 {P}_n^{rs}(L3)+\\ w_3{P}_n^{rs}(L2)+ w_4{P}_n^{cov}+ w_5{E}_n^{res}  , & \text{if $N_C \neq 0$ }\\
			0, & \text{otherwise}
		\end{cases} 
	\end{equation}
	where $w_1 >> w_2 >> w_3 >> w_4 >> w_5$ and $N_C$ is the number of candidate neighbour nodes.
	
	The weights $w_1, w_2, w_3, w_4$, and $w_5$ are chosen based on the significance of the corresponding parameter on data routing. Inter-UAV collision can disrupt the working of the entire network by changing the topology, causing irreparable damage to the UAVs and leading to the loss of information. Accordingly, the $P^{coll}$ receives the highest weightage among other system parameters. $P^{rs}$ is given next higher importance as it indicates the reliability of data transmission. $P^{rs}(L3)$ holds greater significance than  $P^{rs}(L2)$ in ensuring end-to-end data transmission; hence the latter gets higher weightage. $P^{cov}$ offers insights into network characteristics, including topology and wireless channel conditions. While $P^{cov}$ reflects the reliability of links between nodes, $P^{rs}$ delivers real-time feedback on packet reception. Thus, $P^{rs}$ is accorded greater weight than $P^{cov}$. Further energy levels are crucial for UAV operations. However, the remaining energy level influence on routing decisions is subordinate to $P^{cov}$ as robust communication link takes precedence over energy considerations. Without a stable link, data transmission is susceptible to disruptions and failures, even with ample energy reserves. Consequently, to emphasize communication reliability, the weight assigned to $P^{cov}$ is higher than that given to $E^{res}$.  $N_C$ value of 0 indicates the absence of candidate neighbour connections, indicating that the UAV is not connected to the network (fragmented).  A minimum reward of zero for network fragmentation serves as a discouragement to decrease the frequency of such occurrences.
	
	The solution to the formulated  $Q(\lambda)$-learning problem is policy $\pi$, which maps $\pi: \boldsymbol{S} \rightarrow \boldsymbol{A}$ with $\pi(s)$ being the optimal policy taken in state $s\in \boldsymbol{S}$. An optimal policy $\pi^*$  maximizes the expected discounted reward as,
	\begin{equation}
		\pi^* = arg \max_{\pi}\mathbb{E}\bigg[\sum_{n=0}^{\infty}\gamma_nR(s_n,\pi(s_n))\bigg]
	\end{equation}
	where,  $\gamma$, is the discount factor. 
	
	\subsubsection{ $Q(\lambda)$-learning ($Q(\lambda$)) Framework for Multi-hop Routing}
	Q-learning is reward-driven incremental learning that aligns with optimizing routing decisions to enhance successful data transfers. The $Q(\lambda$)  introduces eligibility traces, allowing the algorithm to credit or assign value to the most recent actions and those that led to the current state. Thus enabling $Q(\lambda$) to capture the influence of past actions on the present, providing a more sophisticated approach to learning and decision-making in dynamic environments \cite{Qlambda}. Each element of the Q-matrix is a function of the UAV's state and action and is updated at discrete time instants using the below relation,
	\begin{multline}
		Q^{new}(s_n,a_n) \leftarrow Q^{old}(s_n,a_n) + \beta \big[R_{n+1} +\\ \gamma \max_{a}Q(s_{n+1},a) - Q^{old}(s_n,a_n)\big]e_n(s,a)
	\end{multline}
	where
	\begin{equation}
		e_n(s,a)=\mathcal{I}_{ss_n}\mathcal{I}_{aa_n} + \begin{cases}
			\beta\lambda e_{n-1}(s,a), \hspace{0.9cm}a_n=\pi^*_{n-1}(s_n) \\
			0,\hspace{2.5cm} otherwise
		\end{cases}
	\end{equation}
	where $\mathcal{I}_{xy}$ is an identity indicator function which equals to 1 if $x=y$ and 0, otherwise. 
	
	In (21), the learning rate $(\beta)$  and the discount factor  $(\gamma)$ are dynamic and vary in the range (0,1). $\beta$ controls the Q value update rate, and $\gamma$ determines the relative importance of immediate and future rewards. The dynamic computation of $\beta$  and $\gamma$ are detailed in section IV.D.3. $Q^{old}(s_n,a_n)$ is the old Q value, while $Q^{new}(s_n,a_n)$ is the new updated Q value. The term $\max\limits_{a}{}\ Q(s_{n+1},a)$ estimates the optimal future value, and $R_{n+1}$ is the reward received when transitioning from state $s_n$ to state $s_{n+1}$. In equation (22), the eligibility traces $e(s, a)$, controlled by the learning parameter $\lambda$ (similar to $\beta$), provide a higher update factor for recently revisited state-action pairs $(s, a)$, reinforcing their importance. The eligibility trace is cleared if the previous action $a_n$ is not greedy. Unlike $\beta$, $\lambda$ is static to ensure consistency in memory length so that the UAV relies on a reliable history of experiences.
	
	\subsubsection{Adaptive Learning Rate and Discount Factor}
	In dynamic UAV networks characterized by changing topology and unstable links, fixed learning rates and discount factors are inadequate for effectively adapting to evolving conditions. Thus, the proposed approach introduces adaptability to the learning rate and discount factor, ensuring responsiveness to evolving network conditions.
	
	The dynamic learning rate $\beta_{i}$ of UAV $i$ is defined as
	\begin{equation}
		\beta_{i}= \bigg(\frac{\beta_{max}-\beta_{min}}{1-e^{-P^{cov}(i)}}\bigg)+\beta_{min}
	\end{equation}
	In (23), $\beta_{i}$ is a function of coverage probability ($P^{cov}(i)$). To ensure initial rapid learning and continued exploration as learning advances, we set $\beta_{max}$ = 1 and $\beta_{min}$ = 0.01. As $P^{cov}(i)$ increases, $\beta_{i}$ decreases. Thus allowing lower learning rates in well-connected areas for steady learning and higher learning rates in areas with inadequate coverage and unstable links. 
	
	The dynamic discount factor $\gamma_{i}$ of UAV $i$ is defined as
	\begin{equation}
		\gamma_i=\frac{N_{c}(i)(\gamma_{max}-\gamma_{min})}{M}+\gamma_{min}
	\end{equation}
	
	In (24), $\gamma_i$ is a function involving the ratio of the current number of candidate neighbour nodes ($N_{c}(i)$) to the total count of nodes ($M$). To find an optimal equilibrium between prioritizing immediate rewards and accounting for future considerations to facilitate adaptability to evolving network conditions, we set  $\gamma_{max}$ to 0.9 and $\gamma_{min}$ to 0.1. As $N_c(i)$ increases, $\gamma_i$ also increases, resulting in a higher discount factor in well-connected networks favouring future rewards. Conversely, a lower discount factor in less connected networks to prioritize immediate rewards. 
	
	\begin{algorithm*}[thp]
		\caption{: Improved Q-learning based Multi-hop Routing Algorithm}\label{algo2}
		\begin{multicols}{2}
			\textbf{Input:} UAV initial energy $E$  \\
			\textbf{Output:} Energy efficient reliable multi-hop  data routing path for each UAV-TBS pair $(\pi^*)$\\
			\raggedright{\textbf{Phase 1: Location Acquisition}}\\
			\begin{algorithmic}[1]
				\IF{\textit{time to update node location} arrives}
				\FOR{each UAV $U$ in $U_1$ to $U_M$}
				\STATE obtain \textit{UAV location} information via GPS
				\ENDFOR
				\ENDIF
			\end{algorithmic}
			\raggedright{\textbf{Phase 2: Neighbour Discovery}}\\
			\begin{algorithmic}[1]
				\STATE call \textit{Algorithm 1} for  establishing and updating \textit{NeighbourTable}
			\end{algorithmic}
			\raggedright{\textbf{Phase 3: Parameter Estimation and Initialization}}\\	
			\begin{algorithmic}[1]	
				\STATE Record $E^{res}$ (detailed in \textit{Section IV.B.1})
				\STATE Compute coverage probability (${P}^{cov}$) (17)
				\STATE Compute collision probability (${P}^{coll}$) (18) 	
				\STATE Initialize packet reception status (${P}^{rs}$) (16) = 0 
			\end{algorithmic}
			\raggedright{\textbf{Phase 4: UAV Operational Mode Configuration}}\\
			\begin{algorithmic}[1]	
				\STATE Initialize UAV mode = Neighbour Discovery
			\end{algorithmic}
			\raggedright{\textbf{Phase 5: Routing Decision}}\\
			\begin{algorithmic}[1]
				\WHILE{next-hop node $\neq$ TBS}
				\FOR{each UAV $U$ in $U_1$ to $U_M$}
				\IF{${E}^{res} < {E}_{th}$}
				\STATE Mode = \textit{Charge}
				\IF {${E}^{res} < {E}$}
				\STATE Mode = \textit{Neighbour Discovery}
				\ELSE
				\STATE Mode =\textit{Charge}
				\ENDIF
				\ELSE
				\IF{Mode = \textit{Neighbour Discovery}}
				\STATE Update the \textit{NeighbourTable} (\textit{Algorithm 1})
				\ELSIF{Mode = \textit{Receive}}
				\IF{Data = \textit{HelloMessage}}
				\STATE Update the \textit{NeighbourTable} (\textit{Algorithm 1})
				\ELSIF{Data = acknowledgement (\textit{ACKs})}
				\STATE Update packet reception status (${P}^{rs}$) (16)
				\ELSIF{Data = surveillance information}
				\STATE Receive and push  to transmission queue ($q_t$)
				\ENDIF
				\IF{reception queue is empty ($q_r=0$) \& transmission queue is not empty ($q_t \neq 0$)}
				\STATE Mode is \textit{Transmit}
				\ELSE
				\STATE Mode is \textit{Receive}
				\ENDIF
				\ELSIF{Mode is \textit{Transmit}}
				\STATE Update packet reception status (${P}^{rs}$) (16)
				\STATE Update coverage probability (${P}^{cov}$) (17)
				\STATE Update collision probability (${P}^{coll}$) (18) 
				\STATE Measure reward ($R$) (19)
				\STATE Update Q-matrix.
				\STATE Return optimal next-hop node ($\pi^*$)
				\IF{transmission queue is empty ($q_t=0$)}
				\STATE Mode is \textit{Receive}
				\ELSE
				\STATE Mode is \textit{Transmit}
				\ENDIF
				\ENDIF
				\ENDIF
				\ENDFOR
				\ENDWHILE
			\end{algorithmic}
		\end{multicols}
	\end{algorithm*}
	
	The time complexity analysis of \textit{Algorithm 1: Hello Message Exchange for Topology Construction}, involves two phases. In Phase 1, where each of the $M$ UAVs individually broadcasts \textit{HelloMessage}, the time complexity is O($M$). In Phase 2, the operation involves fetching the positions of $U_i$ (originator UAV) and checking if each $U_i$ is in $N_T$ (\textit{NeighbourTable}). In the worst-case scenario, where each UAV receives messages from all others, and the size of $N_T$ equals $M$, the time complexity is O($M^2$). The overall time complexity of \textit{Algorithm 1} is O($M^2$).
	
	The time complexity analysis of \textit{Algorithm 2: Improved Q-learning based Multi-hop Routing Algorithm}, comprises five phases. Phase 1, obtains $M$ UAV locations with a time complexity of O($M$). Phase 2, calls \textit{Algorithm 1}, with a time complexity of O($M^2$). Phase 3, involves recording energy levels and computing probabilities, resulting in a constant time complexity of O(1). Phase 4, initializes the UAV mode, leading to a constant time complexity of O(1). In Phase 5, the routing decision loop iterates until the next-hop node is the TBS. Within this loop, another loop iterates over each of the $M$ UAVs, performing condition checks, mode updates, and packet handling. Thus, the time complexity of Phase 5 is O($MN$), where $N$ is the number of iterations until the next-hop node is the TBS. The overall time complexity of \textit{Algorithm 2} is O($M^2+MN$).
	
	\section{Results and Discussion}
	This section analyses the performance of the proposed IQMR algorithm through extensive simulation. First, we illustrate the IQMR algorithm decision process. Second, we discuss the behavior of the proposed algorithm for varying system parameters (residual energy, packet reception status, coverage, and collision probabilities). Third, we compare the IQMR algorithm with the existing routing methods, namely Q-FANET\cite{QFANET} and  QMR \cite{QMR} for (i) energy efficiency (rate of decrease in residual energy) and (ii) reliability (number of data packets received at  TBS). Fourth, we evaluate the convergence of the proposed IQMR algorithm for varying exploration rates. 
	\subsection{Simulation Environment}
	UAVs are distributed in a 3D  space, spanning a  radius of 1000 m and a maximum height of 300 m. The UAVs' altitudes vary between 100-300 m, with speeds ranging from 10-30 m/s. Each UAV has a uniform transmission power of 1 W and a radio range of 250 m. In each iteration, the UAVs can transmit directly to TBS or nearby UAVs within the radio range. Each UAV device has a capacity of 207792 (11.1 V, 5200 mAh) Joules of energy. Table II summarizes the detailed parameters considered for simulation. 
	\begin{table}[h]
		\begin{center}
			\caption{List of simulation parameters (\cite{Gupta}\cite{QTAR}\cite{my}\cite{eng-eqs})}
			\begin{tabular}{@{}ll@{}}
				\hline
				\vspace{1pt}
				\textbf {Parameter} & \textbf{Specification}\\ 
				\hline
				network area (cylindrical) & R=1000 m, H=300 m \\
				number of UAVs (${M}$) & 50\\
				number of TBS  & 1\\
				UAV transmission radius & 250 m \\
				UAV transmit power& 1 W\\
				UAV operating altitude $[h_{max}, h_{min}]$ & 100-300 m\\
				UAV speed $[s_{max}, s_{min}]$ & 10-20 m/s \\
				UAV movement direction $[d_{max}, d_{min}]$ & $[-\pi/2, \pi/2]$\\
				UAV vertical pitch $[p_{max}, p_{min}]$ & $[0, \pi/4]$\\
				mobility model & 3D Gauss Markov Mobility  Model\\
				channel propagation model & Nakagami-$m$ small scale fading\\
				${E}$  & 207792 J(11.1V, 5200 mAh)\\
				$E_{th}$& 100 J\\
				$r_0$& 100 m\\
				$\mathcal{E}_{elec}$& 50 nJ\\
				$\mathcal{E}_{amp-fs}$&41 $\mu$J\\
				$\mathcal{E}_{amp-mp}$ & 100 pJ\\
				$\mathcal{E}_{payload}$ & 0.217 kW/kg\\
				$SIR_{th}$  & 0 dB \\
				$P^{cov}_{th}$ & 0.1\\
				$r_{min}$ & 1 m\\
				data packet size & 150 Bytes\\			
				$\xi_x$ = $\xi_y$ & 3m\\
				traffic type& CBR\\
				CBR rate & 2Mbps\\
				MAC protocol & 802.11n\\
				bandwidth & 20 MHz\\
				
				network simulator&MATLAB\\
				\hline
			\end{tabular}
		\end{center}
	\end{table}
	
	\subsection{IQMR routing decision example}
	Fig. 6 shows a simple UAV network topology consisting of a source node (UAV 1), a destination node (TBS), and eight relay nodes (UAVs 2  to 9). Let at time $n$ the source node have a data packet to forward. The source node formulates its neighbour set following the exchange of \textit{HelloMessage}. The candidate neighbour set of UAV 1 includes UAVs 2,3,4 and 9 (as discussed in \textit{Algorithm 1}).
	
	\begin{figure}[]
		\centering
		\subfigure[]{\includegraphics[width=0.49\linewidth]{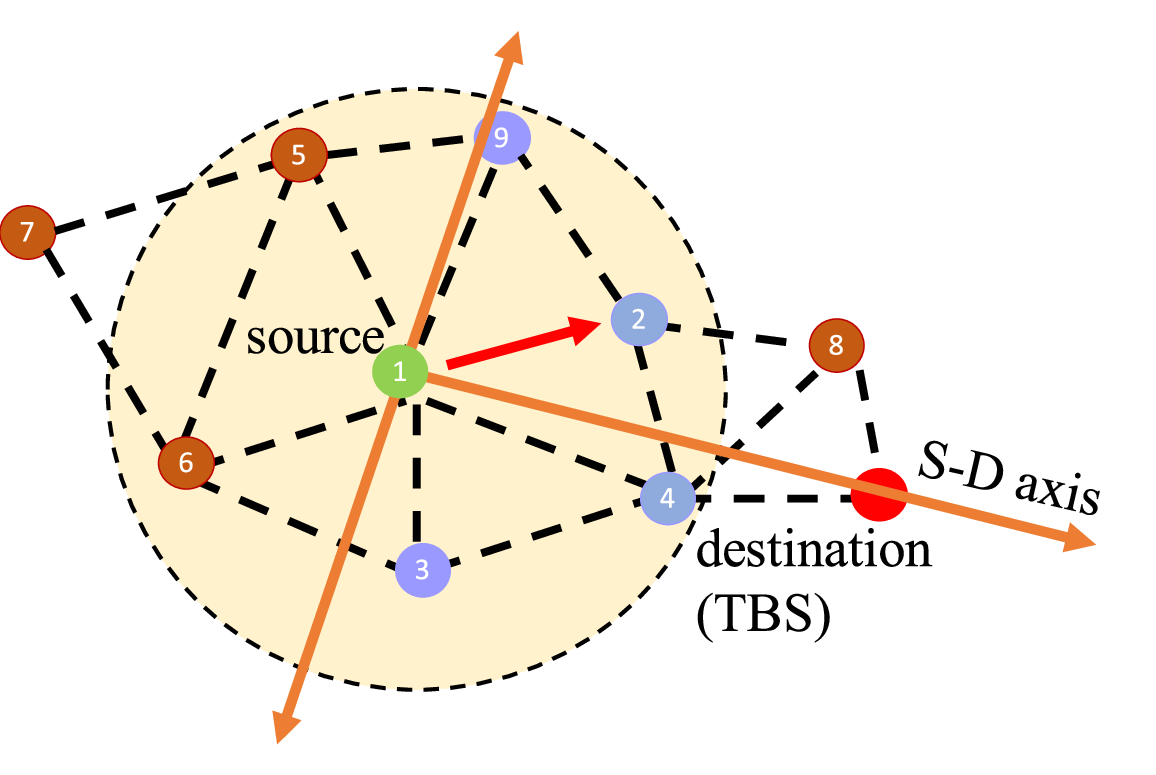}} 
		\subfigure[]{\includegraphics[width=0.49\linewidth]{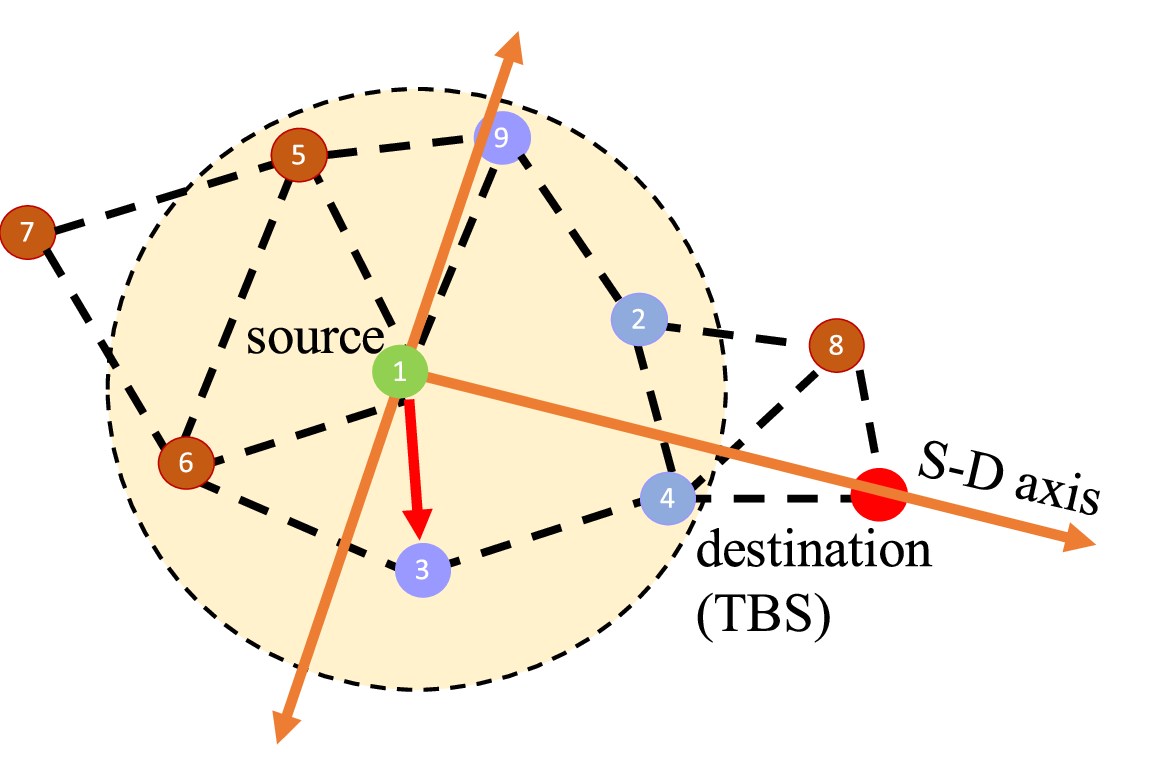}} 
		\caption{Illustration of next-hop node selection for a static network  with a source (UAV 1) and eight relay nodes (UAVs 2 to 9) and destination node (TBS); S-D axis indicating source-destination node axis. Representation of nodes are as green-source node, blue-candidate neighbour nodes, brown- neighbour and non neighbour nodes, red-destination node, grey-candidate nodes that do not satisfy routing constraints.}
		%
	\end{figure} 
	At the time $n$, the Q-values corresponding to the candidate neighbour set [UAV 1, UAV 3, UAV 4, UAV 9] are [0.90, 0.65, 0.85,0.65]. The source node, in general, forwards the packet to the node with the highest Q-value, i.e., UAV 2 (Fig.6a). However, when two candidate neighbours have the same Q-value, then the one less divergent from the source-destination node (S-D) axis is selected as the next forwarder. Assume node 2 and node 4  do not satisfy the constraints (9),(10), or  (11), then the source node does not consider the UAVs 2 and 4 for selection as next-hop nodes. Instead, the source node routes the data to node 3 (Fig.6b). Though the Q-values of nodes 9 and 3 are equal, the latter is selected as the next-hop node since it is comparatively less divergent from the S-D  axis.
	\subsection{Impact of System Parameters on IQMR algorithm}
	This section discusses the impact of varying system parameters such as residual energy, packet reception status, coverage and collision probabilities of the proposed IQMR algorithm. Further, we study the system performance when the UAVs are fragmented from the network. In what follows, in Figs. 7 to 14, the $x$-axis represents the number of episodes of data transmission, and the $y$-axis represents the cumulative reward, except Fig. 9a.
	\subsubsection{Residual Energy}
	The residual energy of the network is the sum of the remaining energies of all the UAVs after each episode of data transmission. Initially, the UAVs are powered with 11.1 V and 5200 mAh battery. The energy expended for data dissemination follows the distance-dependent energy consumption (described in Section IV). This section analyzes the variation of residual energy over the simulated UAV network. Figs. 7 and 8 illustrate the system's cumulative reward behavior when the residual energy of a few UAVs falls below a certain threshold (100 J) simultaneously. Figs. 7a-7d depicts the use case of $20\%$  UAVs (randomly chosen) within the coverage range of the transmitting UAV  with residual energy less than 100 J. When UAVs run out of charge, their ability to complete the surveillance and relay the data is compromised, leading to unstable rewards obtained by the system (indicated as a non-uniform region in the cumulative reward graph). The plateauing in the cumulative reward graph represents the recovery phase, where the system is relatively stable in disseminating information. The system performance is observed over varying learning rates: 0.01 (Fig. 7a), 0.1 (Fig. 7b), 0.5 (Fig. 7c) and 1 (Fig. 7d). The convergence rate of the proposed IQMR algorithm to a stable policy depends on the learning rate, which determines how the algorithm is updated. A higher learning rate implies larger updates, while a lower learning rate leads to smaller updates. Hence, as depicted in the graph, the system with a higher learning rate settles down faster and vice versa.

	\begin{figure}[]
		\centering
		\subfigure[]{\includegraphics[width=0.24\textwidth]{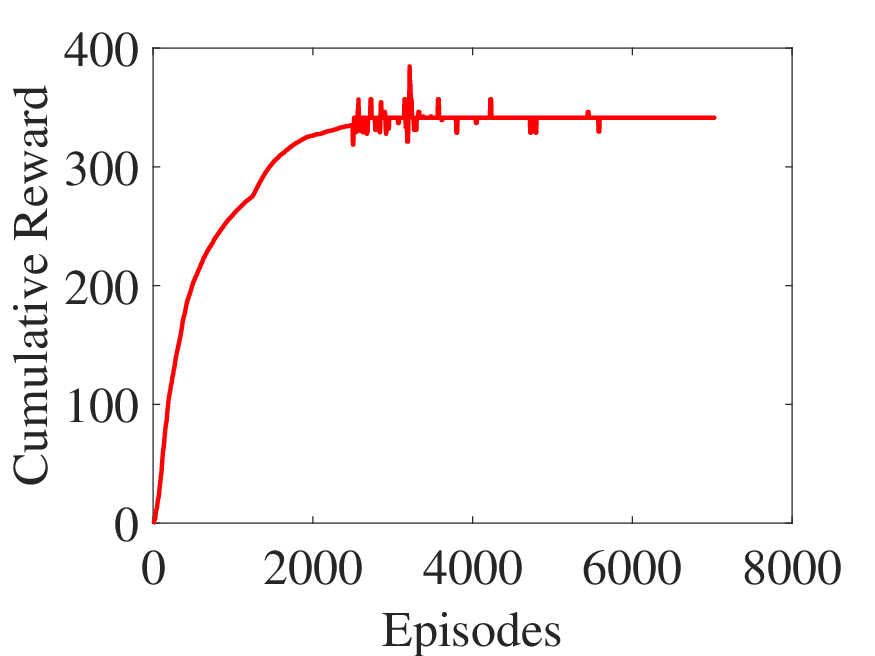}} 
		\subfigure[]{\includegraphics[width=0.24\textwidth]{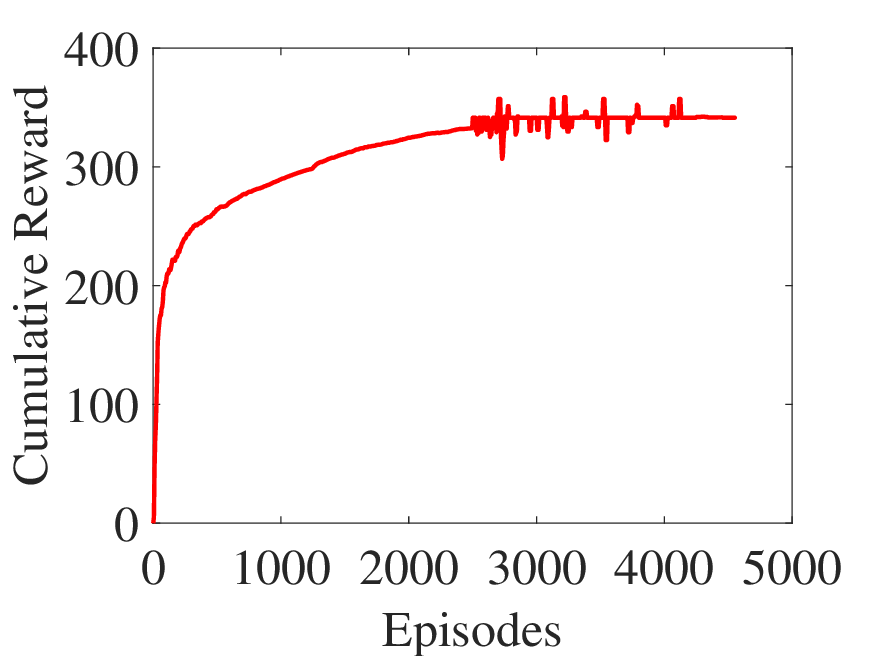}} 
		\subfigure[]{\includegraphics[width=0.24\textwidth]{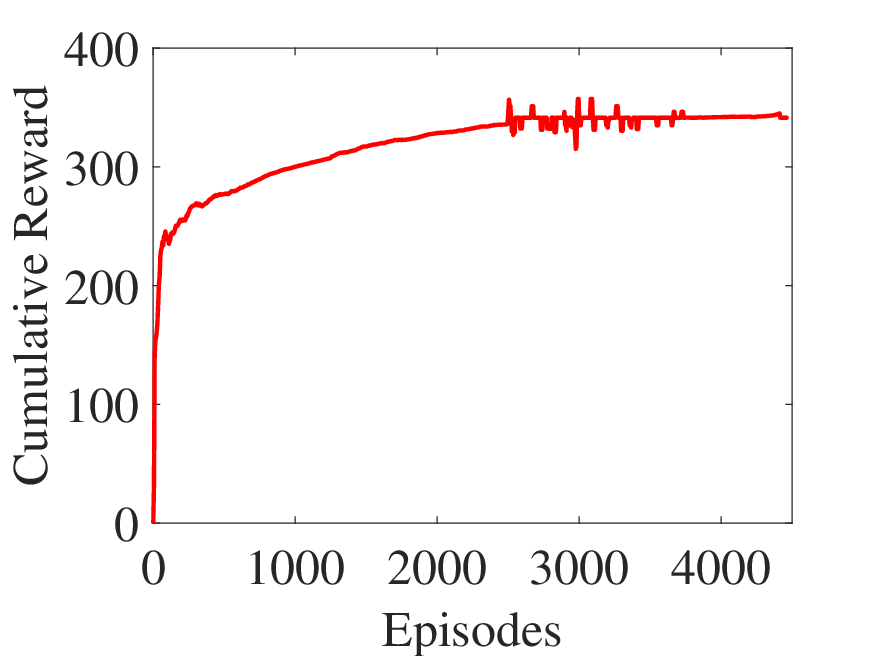}}
		\subfigure[]{\includegraphics[width=0.24\textwidth]{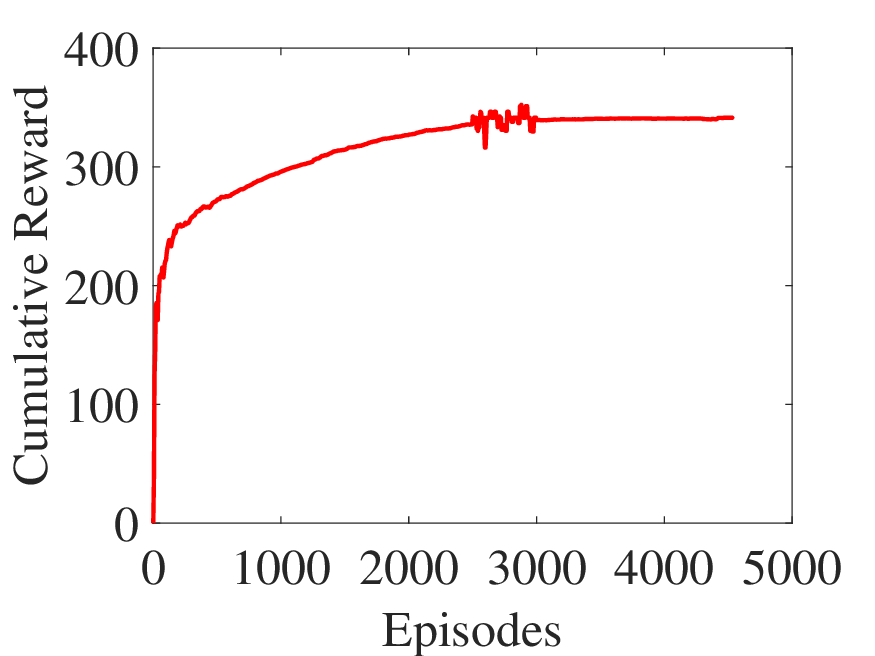}}
		\caption{Impact on cumulative reward  as a function of residual energy for varying learning rates (a) 0.01 (b) 0.1 (c) 0.5 (d) 1.}
		%
	\end{figure} 
	Fig. 8 details the system behavior as a function of residual energy and Q-value. The UAVs are arranged in descending order based on their Q-values and then divided into two sets. Set 1 comprises the top half of UAVs with higher Q-values, while Set 2 includes the remaining half with lower Q-values. This division allows for evaluating the impact of different subsets of UAVs on system performance. The Q-value indicates the suitability of  UAV for forwarding data packets. The departure of UAVs with higher Q-values from the network due to insufficient operational energy ($ {E}^{res}< 100$ J) disrupts the optimal data flow within the network, causing a more pronounced effect on the cumulative reward (Fig. 8a) in comparison to the departure of UAVs with lower Q-values (Fig. 8b).
	
	\begin{figure}[]
		\centering
		\subfigure[]{\includegraphics[width=0.24\textwidth]{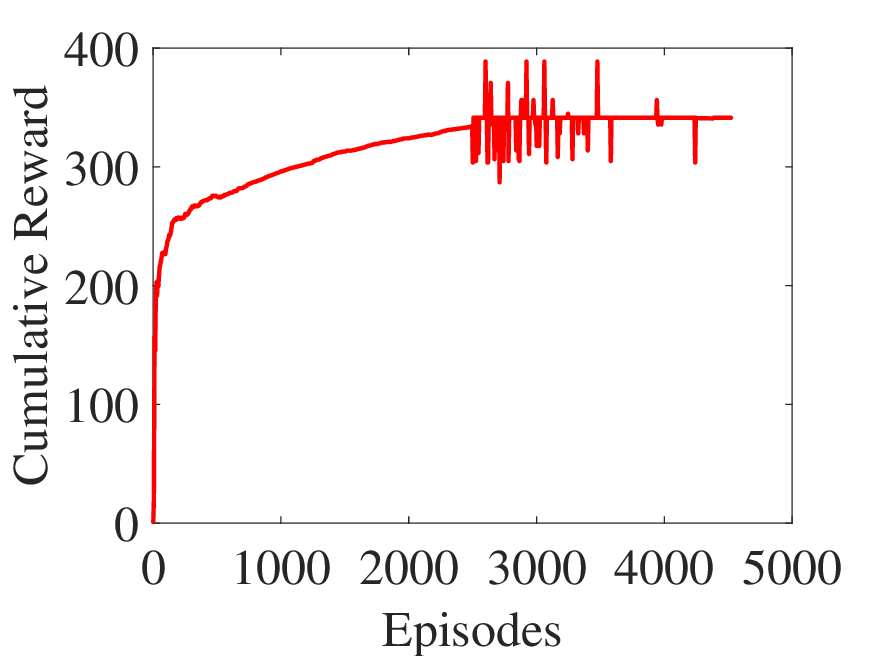}} 
		\subfigure[]{\includegraphics[width=0.24\textwidth]{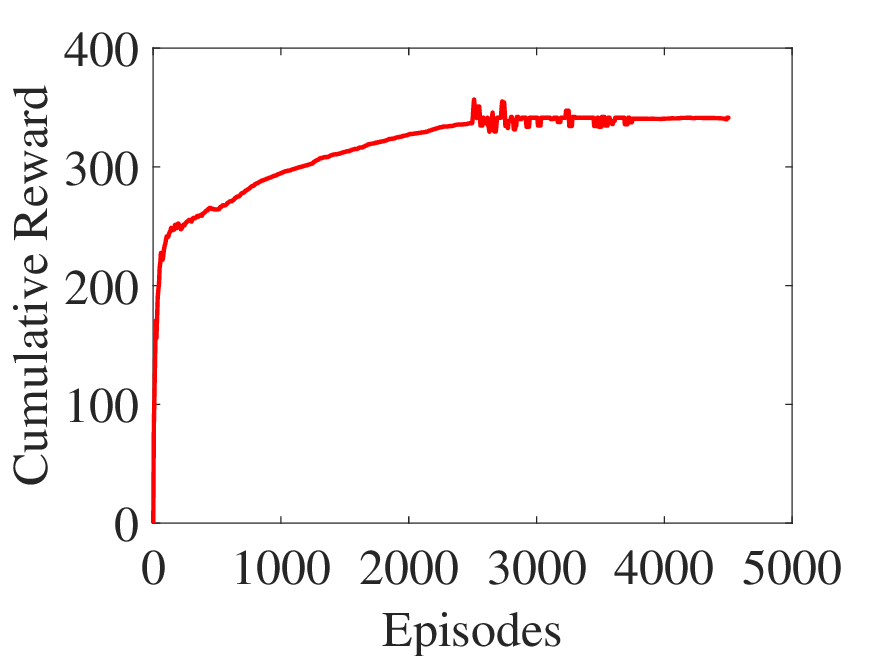}} 
		\caption{Variation of  cumulative reward  as a function of residual energy and Q-values (a) higher Q-values (b)lower Q-values.}
	\end{figure} 
	\subsubsection{Packet Reception Status}
	Fig. 9a details the influence of layers 2 and 3 \textit{ACKs} (\textit{L2} and \textit{L3 ACKs}) on the cumulative reward. \textit{L2 ACKs} are important for next-hop communication while \textit{L3 ACKs} have a broader implication for end-to-end delivery, routing decision and  network-wide operation. Therefore higher cumulative reward is achieved with the reception of \textit{L3 ACKs} than \textit{L2 ACKs}. 
	\subsubsection{Coverage Probability}
	Fig. 9b describes the impact of the SIR threshold ($SIR_{th}$) on the cumulative reward. A higher $SIR_{th}$ value leads to a decrease in ${P}^{cov}$, and vice-versa. A higher threshold reduces coverage as the receiver requires a much stronger signal than the interference level for successful communication, thus limiting the number of UAVs meeting stringent SIR requirements to those close to the transmitter or experiencing low interference levels. On the other hand, a lower threshold increases coverage as the receiver can tolerate a weaker signal relative to the interference level, thus facilitating UAVs located farther away or experiencing higher interference levels to meet the less stringent SIR requirement. Thus, increasing the value of $SIR_{th}$ limits the pool of potential next-hop nodes for data transmission, leading to a decline in the cumulative reward, as illustrated in Fig.8b.
	\subsubsection{Collision Probability}
	Fig. 9c represents the impact of collision cross-sectional area radius ($r$) on the behavior of the cumulative reward of the proposed system. A larger swept area implies a greater likelihood of the UAVs colliding during their trajectories. Thus, an increase in  $r$ negatively impacts the cumulative reward of the system. 
	
	\begin{figure}[]
		\centering
		\subfigure[]{\includegraphics[width=0.24\textwidth]{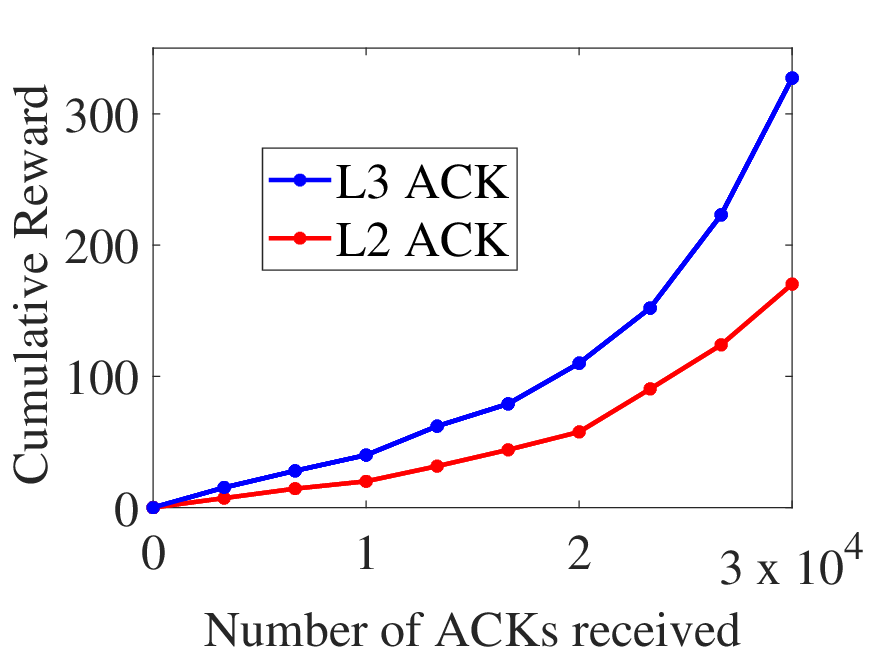}} 
		\subfigure[]{\includegraphics[width=0.24\textwidth]{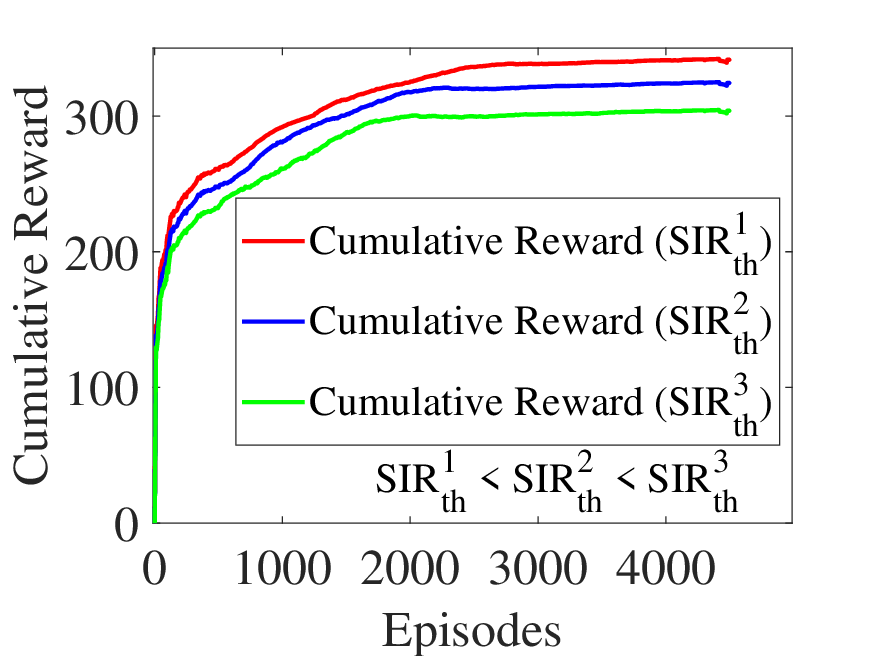}} 
		\subfigure[]{\includegraphics[width=0.24\textwidth]{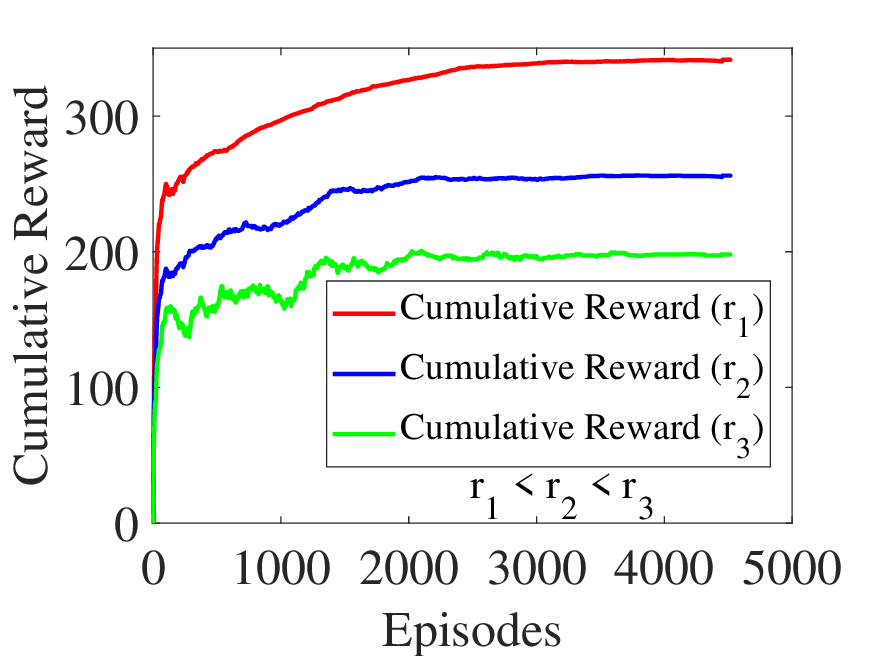}}
		\caption{Impact on cumulative reward for variation in (a) \textit{L2} and \textit{L3} ACKs  (b) SIR (c) inter-UAV distance (\textit{r}) }
		%
	\end{figure} 
	\subsubsection{Network Fragmentation}
	This section analyzes the effect of node fragmentation on the cumulative reward. A proportion of the UAVs are assumed to be fragmented simultaneously and remain out of the network for a fixed duration. The nodes then join the network back together or in batches. Figs. 10 and 11 show cumulative reward variations when higher Q-valued UAVs within the coverage range of transmitting UAV fragments from the network for 400ms and 200ms, respectively, with all (Figs. 10a, 11a) and a quarter (Figs. 10b, 11b) of them rejoining over the next 10ms. Fig. 12 enumerates the impact on the cumulative reward when $20\%$  (randomly chosen) of UAVs within the coverage range of transmitting UAV fragments for 200ms, with all (Fig. 12a) and a quarter (Fig. 12b) of them rejoining over the next 10ms. Fragmentation of higher Q-valued UAVs has significant implications on routing performance, as it decreases the routing efficiency and network resilience and increases communication overhead. Conversely,  simultaneous rejoining (Figs. 10a, 11a, 12a)  causes congestion, collision, and sub-optimal routing. Therefore, gradual rejoining (Figs. 10b, 11b, 12b) facilitates smoother assimilation and faster system stabilization.
	
	\begin{figure}[ht]
		\centering
		\subfigure[]{\includegraphics[width=0.24\textwidth]{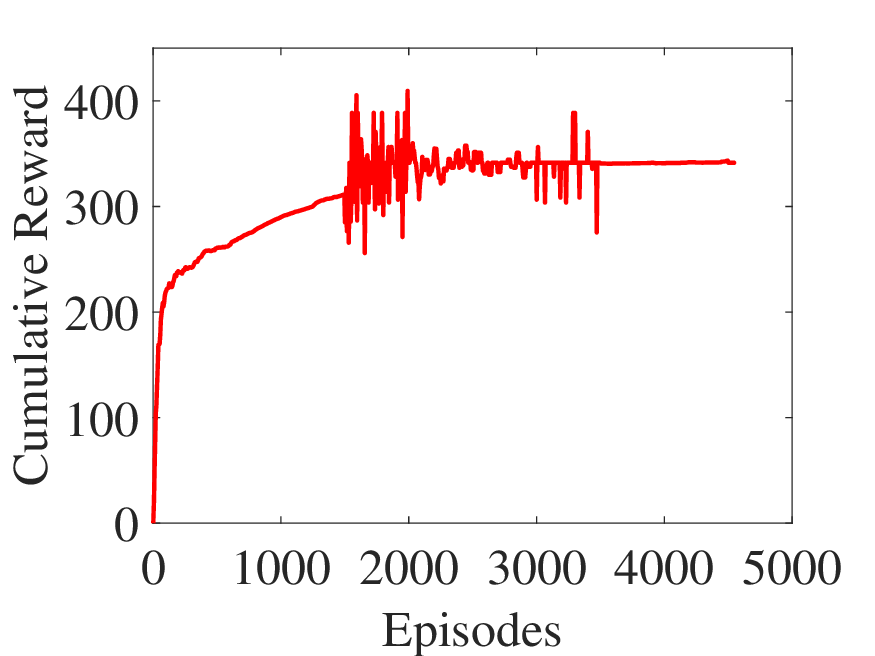}} 
		\subfigure[]{\includegraphics[width=0.24\textwidth]{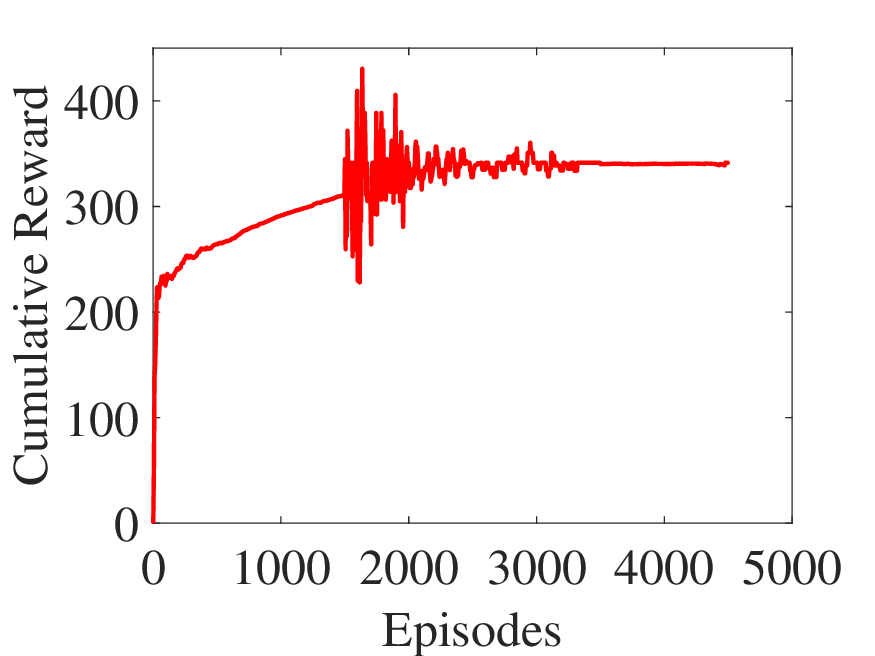}} 
		\caption{Impact on the cumulative reward when higher Q-valued UAVs partition from the network for 400 ms, and  (a) all (b)  quarter join the network back over the next 10 ms.}
	\end{figure} 
	\begin{figure}[]
		\centering
		\subfigure[]{\includegraphics[width=0.24\textwidth]{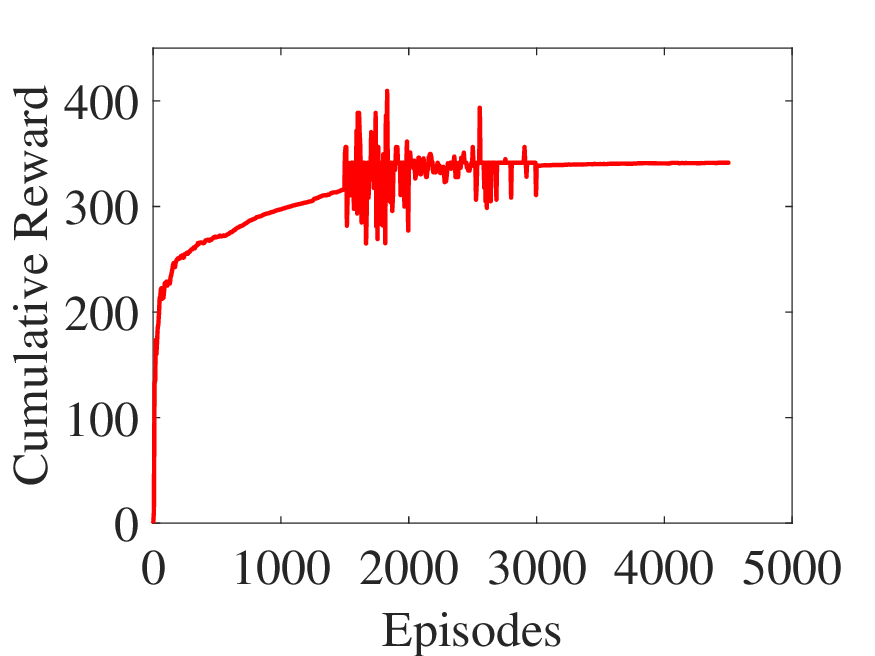}} 
		\subfigure[]{\includegraphics[width=0.24\textwidth]{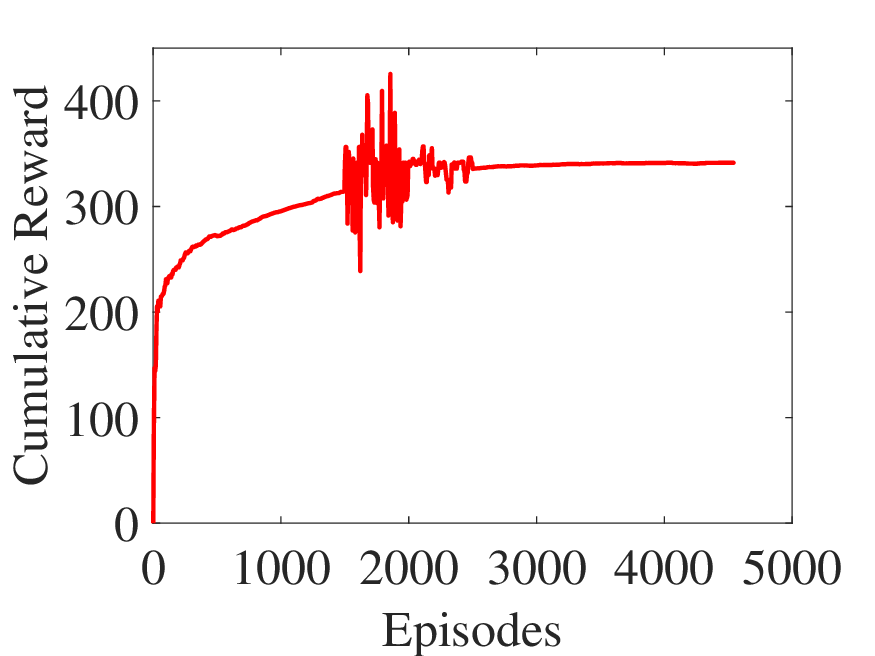}} 
		\caption{Impact on the cumulative reward when higher Q-valued UAVs partition from the network for 200 ms, and  (a) all (b)  quarter join the network back over the next 10 ms. }
	\end{figure} 
	\begin{figure}[]
		\centering
		\subfigure[]{\includegraphics[width=0.24\textwidth]{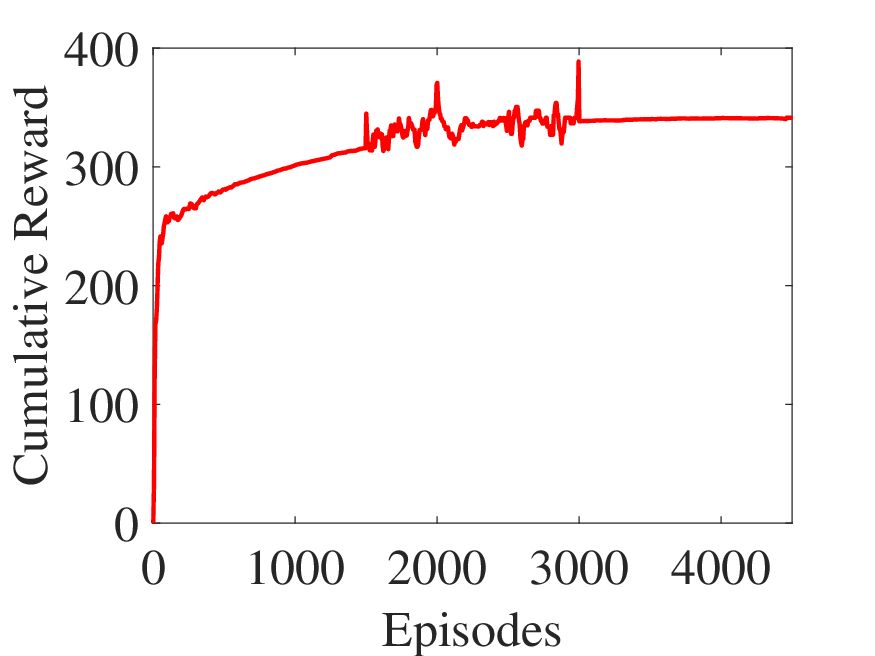}} 
		\subfigure[]{\includegraphics[width=0.24\textwidth]{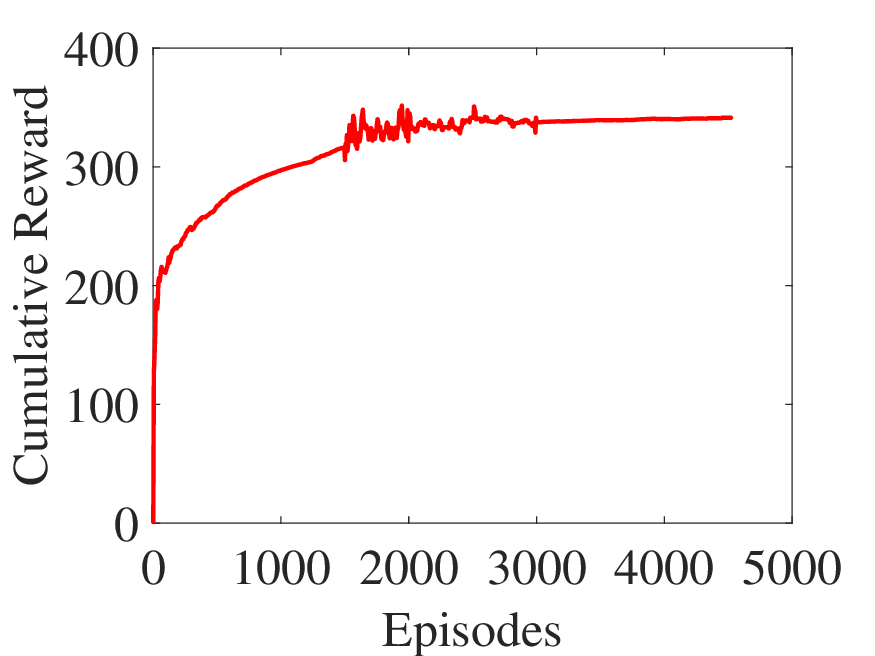}} 
		\caption{Impact on the cumulative reward when $20\%$ UAVs (randomly chosen) partition from the network for 200 ms, and  (a) all (b) quarter join the network back over the next 10 ms.}
	\end{figure}
	
	\subsection{ Performance Comparison: IQMR with Existing Protocols}
	This section compares and analyses the performance of the proposed IQMR algorithm with the existing routing methods Q-FANET and QMR. 
	\subsubsection{IQMR Energy Efficiency Analysis}
	As depicted in Fig. 13a, the IQMR algorithm demonstrates superior energy consumption compared to Q-FANET and QMR. For the  Q-FANET protocol, the residual energy reduces to 0 after 5000 episodes. In contrast, by the 8000 episode, QMR retains only 4.07\% of its initial energy, whereas IQMR retains 36.34\%. Thus, IQMR attains 32.27\% and 36.35\%,  higher energy consumption efficiency than QMR and QFANET. IQMR's adeptness in selecting energy-efficient UAVs as next-hop nodes ensures balanced energy consumption across the network. Furthermore, IQMR's propensity for selecting higher-quality links translates to reduced re-transmissions and elevated energy efficiency.
	\begin{figure}[]
		\centering
		\subfigure[]{\includegraphics[width=0.24\textwidth]{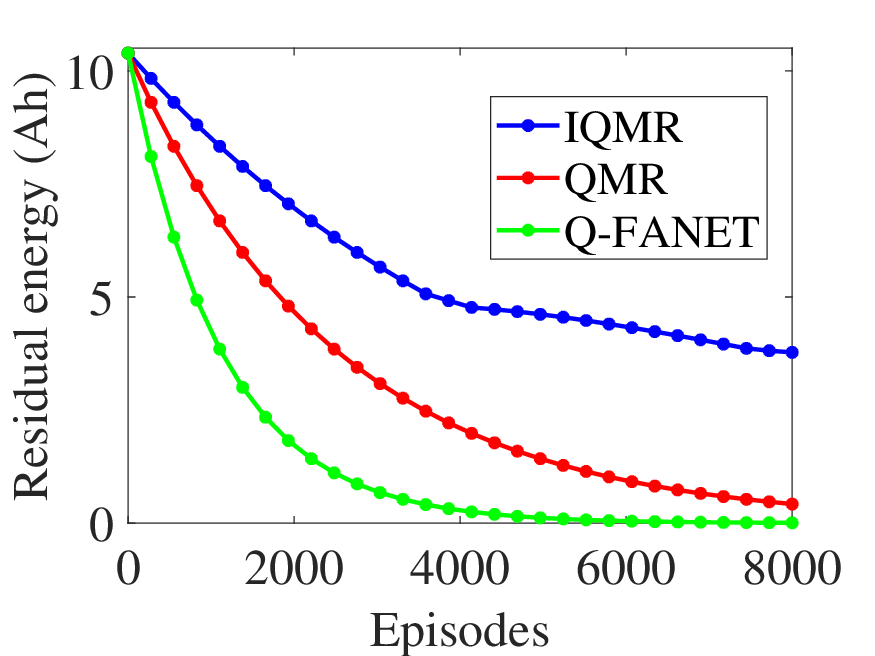}} 
		\subfigure[]{\includegraphics[width=0.24\textwidth]{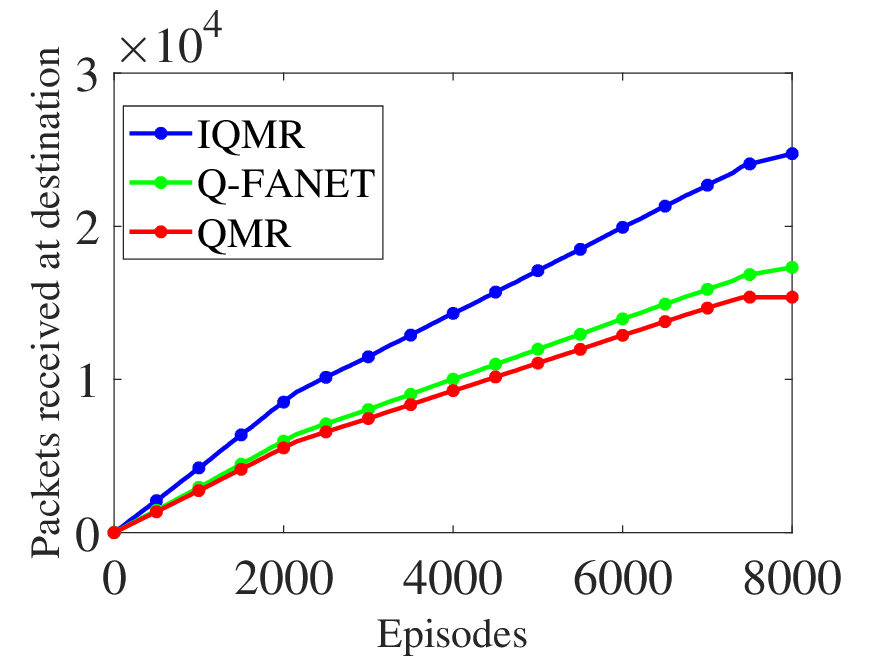}} 
		\caption{Illustration of (a) residual energy and (b) data throughput over varying episodes of data transmission.}
	\end{figure} 
	\subsubsection{IQMR Data Throughput Analysis}
	Fig. 13b illustrates the data throughput quantifying the number of successfully transmitted packets to the TBS. The total number of transmitted packets in the network is  $3\times10^4$. At the 8000-episode mark, IQMR successfully sends about $2.5\times10^4$ packets to the TBS. At the same time, QMR and  Q-FANET manage approximately $1.8\times10^4$ and $1.5\times10^4$  successful data packet transfers to the TBS. Thus, IQMR achieves 25.19\% and 32.05\%   higher data throughput than QMR and Q-FANET, respectively. IQMR achieves substantial data throughput improvement by selecting optimal next-hop nodes, reducing data loss through collision avoidance, and successfully assessing data reception.
	\subsection{IQMR Convergence}
	The proposed IQMR algorithm's convergence is assessed under varying exploration rates: $\epsilon$ values of 0.1, 0.5, and 0.9. As depicted in Fig. 14, the algorithm stabilizes in fewer than 500 episodes and gradually converges, demonstrating its stability and reliability over time, ultimately determining its suitability and effectiveness in real-world network environments. The cumulative reward decreases as $\epsilon$ increases, as with lower $\epsilon$, the UAVs tend to select the next-hop node based on the maximum Q-value, leading to better decisions regarding successful data transmission. Thus, this result underscores the importance of choosing an appropriate exploration rate to ensure adequate training and decision-making.
	\begin{figure}[]
		\centering
		{\includegraphics[width=0.24\textwidth]{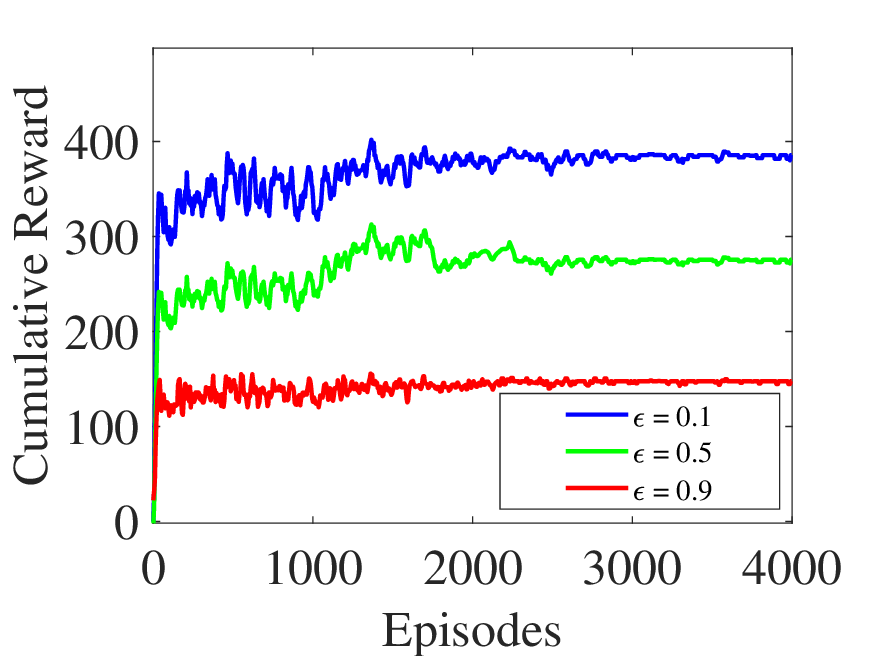}} 
		\caption{Convergence of IQMR for varying exploration rates ($\epsilon$ = 0.1, 0.5 and 0.9)}
	\end{figure} 
	
	\section{Conclusion}
	UAVs are poised to revolutionize 5G and beyond networks, offering unprecedented opportunities for communication and sensing. This study presents a novel IQMR algorithm for multi-hop data routing in dynamic UAV networks. IQMR algorithm's significance lies in its ability to address challenges in changing environments such as search and rescue missions, environmental monitoring, traffic management, and urban planning, where real-time adaptation, resilience to disruptions and efficient resource utilization are paramount. Compared to traditional approaches, IQMR offers a practical solution that ensures energy efficiency, connectivity, and inter-UAV collision prevention while facilitating network restoration without requiring prior UAV path planning. The study analyses how IQMR responds to fluctuating system parameters amid intermittent disruptions in the network environment, demonstrating stabilization within a reasonable time and converging in fewer than 500 episodes, emphasizing its stability. A comparison of IQMR with QMR and Q-FANET protocols shows IQMR's superior energy efficiency and data throughput. IQMR achieves energy consumption efficiency improvements of 32.27\% and 36.35\% over QMR and Q-FANET, along with significantly higher data throughput enhancements of 25.19\%, 32.05\%, over Q-FANET and QMR, respectively.
	
	Future extensions to the IQMR algorithm include integrating mobility prediction, addressing scalability and large-scale deployment, and integrating edge computing capabilities.

	\section*{Acknowledgment}
	The authors thank the Department of Science and Technology (DST), Government of India , for funding this work under the project Advanced Communication System included in National Mission on Interdisciplinary Cyber Physical Systems.

\end{document}